%% file: main.tex
\documentclass[sigconf,10pt]{acmart}
\renewcommand\footnotetextcopyrightpermission[1]{}

\usepackage{graphicx}
\usepackage{textcomp}
\usepackage{hyperref}
\usepackage{xcolor}
\usepackage{balance}
\usepackage{subcaption}
\usepackage{multirow}
\usepackage{booktabs}
\hypersetup{
    colorlinks=true, 
    linkcolor=black, 
    filecolor=black,
    urlcolor=black,
    citecolor=black, 
}

\usepackage{url}
\usepackage{soul}
\usepackage{makecell}
\usepackage{float}
\usepackage{amsmath}
\usepackage{algorithm}
\usepackage{algpseudocode}
\usepackage{algpseudocodex}
\usepackage{amsthm} 
\usepackage{lipsum}
\usepackage{gensymb}
\usepackage{enumitem}
\usepackage{bbm}

\newcommand\blfootnote[1]{%
  \begingroup
  \renewcommand\thefootnote{}\footnote{#1}%
  \addtocounter{footnote}{-1}%
  \endgroup
}

\newcounter{observation}
\renewcommand{\theobservation}{\arabic{observation}}

\algrenewcommand\algorithmicindent{1.0em}

\title{OrbitTransit: Traffic Delivery and Diffusion for Earth Observation via Satellite Mobility}

\author{Haoyuan Zhao}
\affiliation{
  \institution{Simon Fraser University}
  \city{Burnaby}
  \country{Canada}
}
\email{hza127@sfu.ca}

\author{Long Chen}
\affiliation{
  \institution{Simon Fraser University}
  \city{Burnaby}
  \country{Canada}
}
\email{longchen.cs@ieee.org}

\author{Yi Ching Chou}
\affiliation{
  \institution{Simon Fraser University}
  \city{Burnaby}
  \country{Canada}
}
\email{ycchou@sfu.ca}

\author{Hao Fang}
\affiliation{
  \institution{Simon Fraser University}
  \city{Burnaby}
  \country{Canada}
}
\email{fanghaof@sfu.ca}

\author{Jiangchuan Liu}
\affiliation{
  \institution{Simon Fraser University}
  \city{Burnaby}
  \country{Canada}
}
\email{jcliu@sfu.ca}

\begin{document}
\begin{abstract}

The emerging demand for Earth observation (EO) to address environmental challenges has driven unprecedented growth in its primary carrier, Low Earth Orbit satellites, in recent years. Ground stations (GSs), the egress points of these networks, are congested due to the massive volume of EO traffic, and their deployment is constrained by geographic, political, and budgetary factors. Although inter-satellite links (ISLs) can partially relieve this congestion by forwarding traffic to alternative GSs, existing ISL-based approaches can hardly address traffic contention caused by biased GS distribution and may also raise sustainability concerns due to prolonged ISL paths. In this paper, we propose OrbitTransit, a pickup-carry-offload (PCO) approach that leverages satellite mobility for data \textit{delivery} and integrates ISLs for traffic \textit{diffusion} to alleviate the resource contention inherent in PCO delivery. The proposed orbit-as-node framework and contention-avoidant delivery jointly determine the optimal hybrid PCO-ISL path, minimizing energy consumption and balancing GS traffic. Extensive experiments show that OrbitTransit reduces battery consumption by $47.16\%$, decreases task failures by $1.09\times$, and improves GS load balancing compared with state-of-the-art GS selection and routing algorithms.

\end{abstract}
\maketitle
\blfootnote{Haoyuan Zhao and Long Chen contributed equally to this work.}
\blfootnote{To appear in the Proceedings of ACM MobiSys 2026.}

\input{sections/1-intro}
\input{sections/2-background}

\input{sections/2.1-insight_and_takeaways}

\input{sections/3-formulation}
\input{sections/4-method}
\input{sections/5-system_design}

\input{sections/6-evaluation}
\input{sections/7-related_work}
\input{sections/8-conclusion}

\bibliographystyle{unsrt}
\bibliography{reference.bib}

\newpage
\input{sections/9-Appendix}

\end{document}

%% file: sections/1-intro.tex
\begin{figure}[t]
     \includegraphics[width=\linewidth]{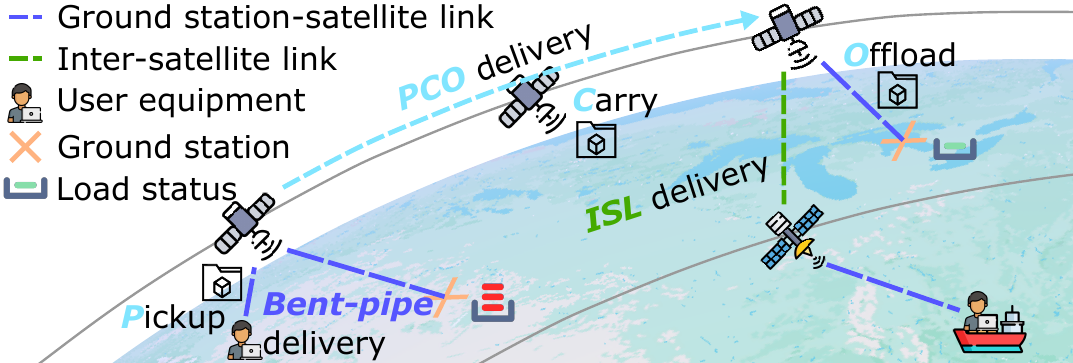}
     \caption{Overview of the typical backhaul path in Low Earth Orbit satellite networks.}
     \label{fig:intro_overview}
\end{figure}

\section{Introduction}

As environmental degradation worsens and becomes more frequent~\cite{nbcnews2024californiafires}, Earth Observation (EO) has become critical for monitoring and research. Recently, Low Earth Orbit (LEO) satellites have emerged as a promising space data delivery medium and observation platform, offering download bandwidths of up to 200 Mbps~\cite{ma2023network, mohan2024multifaceted} and fine-grain sensing resolution of 30~m~\cite{nasa_landsat8}, with leading operators such as Starlink, Kuiper, and OneWeb. However, a single EO mission can generate tens of terabytes of data per day~\cite{nesdis_jpss_cloud_2021, nesdis_viirs_2025}, including high-resolution spectral data and images, which must be offloaded to ground infrastructure for processing and analysis. With the increasing demand for EO tasks and the growing number of sensing satellites, the volume of global EO data can easily accumulate to the petabyte scale. Accordingly, the ground stations (GS), the egress points of LEO satellite networks (LSNs), have gradually become congestion hotspots, noticed by both the user community and researchers~\cite{li2024stable, tao2023transmitting, hughes2023onewebgateways}. 

As shown in Figure~\ref{fig:intro_overview}, typical LSN backhaul relies on nearby GSs to ``bent-pipe''~\cite{ma2023network, tao2023transmitting, zhao2023realtime} data through GS-satellite link (GSL), but GS deployment is constrained by geographic, political, and budgetary, leading to a scarcity of GSs in remote regions. To address this, inter-satellite links (ISLs) enable satellite-to-satellite laser communication, allowing remote users (e.g., ships and islands) to reach distant GSs via multi-hop backhaul. For example, recently introduced LEO observation platforms, such as NASA’s Near Space Network~\cite{NASA_NearSpaceNetwork, NASA_WhatIsNearSpaceNetwork}, are capable of both ISL communication and EO missions. Accordingly, recent studies propose GS selection~\cite{chou2025commercial, tao2023transmitting} and ISL routing~\cite{vasisht2021l2d2, li2021processing, lai2023achieving, cao2023satcp, liu2024orbit, chen2025demand, chen2021time} to alleviate GS congestion, reduce delays, and minimize energy consumption.

Moreover, leveraging the high-speed orbital motion of LEO satellites ($\approx 7.8$~km/s), EO data can be \textit{picked up}, \textit{carried}, and \textit{offloaded} at downstream GSs without relying on ISLs, a strategy known as PCO delivery in Figure~\ref{fig:intro_overview}. PCO significantly reduces the energy consumed for data propagation, since satellites traverse their orbits ballistically without additional energy input, whereas a typical laser transmitter can consume up to 35~Watt~\cite{TESAT2024}. Although the 90-minute orbital period of LEO satellites can substantially delay delivery compared with ISLs, EO data are generally intended for long-term monitoring and analysis, with collection cycles spanning hours or even days~\cite{nasa_earthdata_2024}. Thus, such delivery delays remain acceptable.

However, through subsequent trace-driven simulation and analysis, we observe that these solutions cannot fundamentally resolve GS congestion for two challenges: (i)~\textbf{Biased GS distribution.} Traditional GS selection~\cite{chou2025commercial} and ISL-based routing~\cite{vasisht2021l2d2, lai2023achieving, liu2024orbit, chen2025demand, chen2021time} cannot fundamentally overcome traffic hotspots caused by the scarcity of GS resources, which stems from the uneven deployment of GSs and their strong coupling to population distribution. (ii)~\textbf{Orbital resource contention.} Although PCO delivery reduces ISL usage by leveraging satellite mobility, existing methods~\cite{tao2023transmitting, spacebelt2025, huang2018envisioned} inevitably underutilize orbital resources when offloading to target GSs, leading to depleted storage and energy, delaying delivery, and raising battery sustainability concerns.

In this paper, we propose OrbitTransit, a hybrid PCO-ISL \textit{delivery} system for EO data that mitigates GS congestion and orbital resource contention through traffic \textit{diffusion}. Specifically, to simplify the dynamic nature inherent in LSN modeling, we introduce the Orbit-as-Node (OAN) framework, which leverages static orbital parameters to reduce the scale of the topology and abstract GSL and ISL updates without compromising optimality. For the challenge (i), we design an OAN-based GS traffic diffusion mechanism that determines the optimal GS destination for each flow and mitigates distribution bias by balancing traffic across orbits through ISL delivery. To address challenge (ii), we propose a contention-avoidant delivery scheme that combines PCO with minimal ISL usage to offload data to target GSs, avoiding orbital resource contention and prioritizing real-time delivery tasks.

To evaluate OrbitTransit, we build a control plane prototype based on real-world LSN topologies, satellite configurations, and ground infrastructure documentation. The control plane records the position and resource status of each satellite and GS, and updates ISL and GSL connectivity according to existing work and documentation~\cite{mohan2024multifaceted, spacex_services_fcc, FCC2025Starlink40Antennas}. We compare OrbitTransit against four state-of-the-art GS selection and space routing solutions. Extensive experimental results show that OrbitTransit reduces battery consumption by 47.16\%, achieves $1.09\times$ fewer delivery failures, and limits the maximum GS queueing delay to 4.75~ms. Our main contributions are summarized as follows.

\begin{itemize}[left=0pt]
\item We conduct a trace-driven study of existing GS selection and routing methods, and find that their limitations primarily stem from biased ground station deployment and underutilized orbital resources, then explore potential solutions to fundamentally address these issues (\S\ref{sec:background}, \S\ref{sec:analysis_and_insights}).
\item Based on these insights, we propose the OrbitTransit, a sustainable and contention-aware traffic diffusion framework that leverages hybrid PCO-ISL delivery using the orbit-as-node framework. We formulate and model the OrbitTransit system and demonstrate how it addresses the aforementioned issues (\S\ref{sec:modeling}, \S\ref{sec:method}).
\item We build the LSN control plane and implement the OrbitTransit system. Extensive experiments demonstrate the effectiveness of OrbitTransit in balancing GS load, extending satellite lifespan, and achieving higher delivery capacity compared to four state-of-the-art solutions across various constellation setups (\S\ref{sec:evalution}).
\end{itemize}

%% file: sections/2-background.tex
\section{Background of LSNs}
\label{sec:background}

\textbf{Ground infrastructure in LSNs.} Ground stations and points of presence (PoPs) are the primary bridge between LSN traffic and terrestrial networks. As shown in Figure~\ref{fig:intro_overview}, user equipment (UE) sends packets to satellites, which forward the data to nearby GSs or neighboring satellites via ISL, depending on GS availability. GSs receive packets and forward them to PoPs over fiber optic links for high-speed backhaul. Inbound traffic follows a similar reverse path: packets sent from servers pass through the PoP and GS and reach the UE.

\textbf{Backhaul strategies in LSNs.} The bent-pipe strategy, where a satellite simultaneously visible to both the UE and a GS directly forwards UE packets to the GS, is the dominant offloading approach in today’s LSNs~\cite{ma2023network, tao2023transmitting, mohan2024multifaceted}. This method proved feasible in early Starlink deployments, offering short routing paths and low energy consumption. However, a typical Starlink GS requires an unobstructed minimum elevation angle of $25\degree$ to establish a GSL, yielding a coverage circle of roughly $2{,}500$~km in diameter~\cite{mohan2024multifaceted}. As a result, the bent-pipe strategy quickly falls short for users outside GS coverage.

\textbf{Roles of ISLs in LSNs.} Apart from underdeveloped regions, users on islands and cargo ships also lack access to GSs. To serve LSN users not covered by nearby GSs, ISLs enable communication with neighboring satellites, allowing traffic to be forwarded across the constellation and offloaded to distant GSs for backhaul. Since version V1.5, all new Starlink satellites have carried ISL transmitters, with the adoption ratio already reaching 80\% and steadily rising~\cite{mcdowell2025starlinkstats}. Other LSN providers, such as OneWeb and Kuiper, are also actively deploying ISL technology.

Although ISLs can theoretically support Gbps-level transmission rates~\cite{zhai2024seco, cao2023computing}, their actual performance can be severely affected by the chaotic space environment~\cite{ma2023network, yue2023low}, routing updates~\cite{izhikevich2024democratizing, mohan2024multifaceted}, and traffic bursts during Internet peak hours~\cite{zhao2023realtime, fang2024robust}. Therefore, it remains essential to limit ISL usage and reserve bandwidth and energy~\cite{vasisht2021l2d2, lai2023achieving, liu2024orbit, chen2025demand, chen2021time} for real-time applications~\cite{zhao2025baroc, zhang2024starstream, zhao2024low, fang2024robust}, especially given that EO data are generally delay-tolerant and massive in volume.

\textbf{Energy sustainability in LSNs.} Unlike other edge devices, satellite batteries are difficult, if not impossible, to replace. Moreover, half of a LEO satellite's orbital period occurs on the dark side, without solar panel charging, while energy consumption for basic operations, communications, and onboard processing remains continuous. Consequently, efficient energy management, such as limiting prolonged or intensive ISL communications and processing~\cite{chen2025demand, chou2025commercial, liu2024orbit}, is critical for extending battery lifespan and ensuring the sustainability of LSN constellations.

\textbf{Integrated EO-communication systems}. Driven by growing demand for low-latency communication and on-orbit observation, integrated platforms combining communication and sensing functions are emerging. Representative examples include Starshield~\cite{SpaceX_Starshield}, launched in April 2025, NASA’s Near Space Network (NSN)~\cite{NASA_NearSpaceNetwork, NASA_WhatIsNearSpaceNetwork}, introduced in early 2025, and traditional EO leaders such as Planet Labs~\cite{NASA_CommercialSpaceComms}, increasingly moving toward integrated architectures. For example, NSN supports optical relay and inter-satellite communication~\cite{NASA_NSN_UsersGuide}, with data rates up to $5$~Gbps and a long-term target of $\geq 100$~Gbps, enabling navigation, tracking, telemetry, and remote sensing data relay. Although Starshield’s detailed specifications remain undisclosed, public reports and official statements suggest that it also supports ISL communication, like standard Starlink satellites, and can accommodate multiple EO-related tasks~\cite{Misturado2024, NewSpaceEconomy2024, FraunhoferFHR2022}. These developments indicate that integrated EO-communication systems will become increasingly important.

%% file: sections/2.1-insight_and_takeaways.tex
\section{Measurement and Insights}

\begin{figure}[t]
     \centering
     \includegraphics[width=\linewidth]{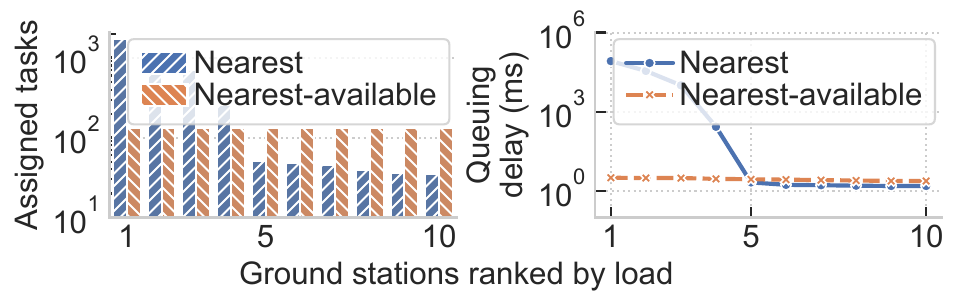}
     \caption{Number of assigned tasks and queuing delays for the top-10 loaded ground stations.}
     \label{fig:gs_status_hot_vs_nearest}
\end{figure}

In this section, we conduct a trace-driven study of existing methods for GS selection and routing. We analyze the root causes of their failures from the perspectives of ground infrastructure, orbital resources, and spatiotemporal dynamics, deriving insights that point to potential solutions.

\textbf{Existing GS selection algorithms.} To address GS congestion issues, two general methods are commonly used to select appropriate destination GSs. (i)~Nearest selection: The satellite selects the nearest GS to offload traffic, behaving like the bent-pipe strategy when the nearest GS is within communication range. This approach has been observed in the early stages through Starlink traffic analyses~\cite{ma2023network, tao2023transmitting, zhao2023realtime}. (ii)~Nearest-available selection: As the user base grows and ISLs become available between most satellites, the nearest GS often becomes suboptimal due to regional policies, visibility constraints, and load conditions \cite{izhikevich2024democratizing}. In such cases, traffic is rerouted to more distant GSs to optimize overall constellation performance and balance load, a behavior also observed in current Starlink topologies~\cite{mohan2024multifaceted}.

\textbf{Existing space routing algorithms.} Once the destination GS is selected, there are two major approaches to deliver the data. (i)~ISL-based routing: The satellite routes traffic to the destination GS via ISLs, with the path determined by factors such as delay, bandwidth, satellite resource, and energy consumption~\cite{yang2016towards, liu2024orbit, chen2022delay, lai2023achieving, pan2025stableroute}. In this paper, we categorize the bent-pipe strategy as an edge case of ISL-based routing with zero ISL connections. (ii)~PCO delivery: The satellite picks up data from the user, carries it, and offloads it to the destination GS through its mobility~\cite{tao2023transmitting}. In this way, it saves energy otherwise consumed by ISL communication.

\textbf{Simulation setup.} The constellation is configured using real-world two-line element data~\cite{celestrak2025} from Starlink and GS infrastructure extracted from FCC filings~\cite{FCC2025Starlink40Antennas}. The EO traffic is generated based on the volume and deadlines of real-world EO missions. The key evaluation metrics include GS load balance, routing efficiency, and orbital resource utilization to assess the performance of existing methods\footnote{For a detailed description of the experimental setup, see \S\ref{sec:evalution}.}.

\label{sec:analysis_and_insights}

\subsection{Lessons from ISL-based Algorithms}

\begin{figure}[t]
     \centering
     \includegraphics[width=\linewidth]{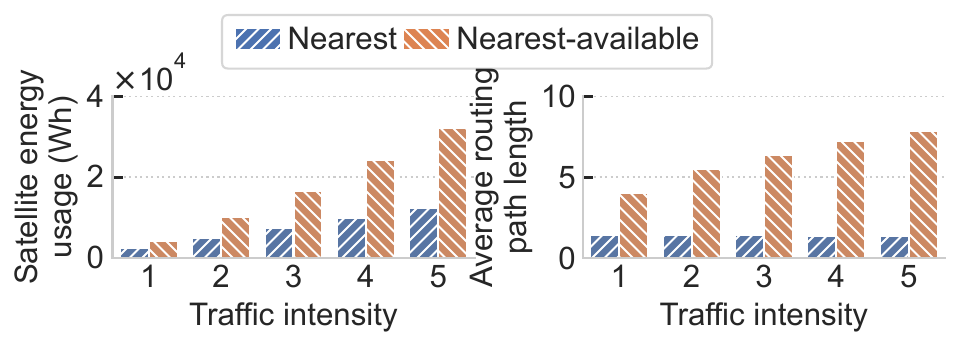}
     \caption{Satellite onboard battery consumption and routing path length under two GS selection algorithms.}
     \label{fig:moti_near_vs_nearavail}
\end{figure}

\textbf{Performance of ISL-based algorithms.} We first evaluate ISL-based routing with two GS selection algorithms, as shown in Figure~\ref{fig:gs_status_hot_vs_nearest}. Under nearest selection, the top four loaded GSs become congested due to excessive assigned tasks, resulting in queuing delays of up to $10^6$~ms. This effectively renders these GSs unresponsive, while 97\% of the remaining GSs operate below 50\% load, leading to low overall utilization. In contrast, nearest-available strategically offloads data to farther GSs, achieving a more balanced load distribution and much lower queuing delays, with no GS overloaded and a maximum delay of 3.42~ms. However, its extended ISL paths often deplete onboard batteries, raising sustainability concerns, as shown in Figure~\ref{fig:moti_near_vs_nearavail}. Energy consumption and path length also increase with higher traffic intensity, since nearby GSs become fully loaded and satellites must offload to more distant GSs.

\begin{figure}[t]
     \centering
     \includegraphics[width=0.9\linewidth]{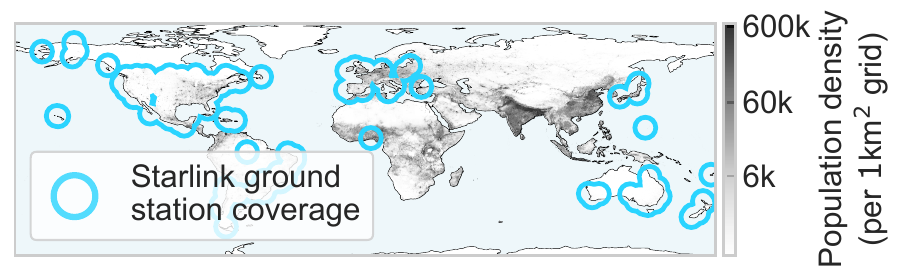}
     \caption{The distribution of ground station coverage and population densities.}
     \label{fig:population_and_gs}
\end{figure}

\begin{figure}[t]
    \centering
    \begin{minipage}[t]{0.23\textwidth}
        \centering
        \includegraphics[width=\textwidth]{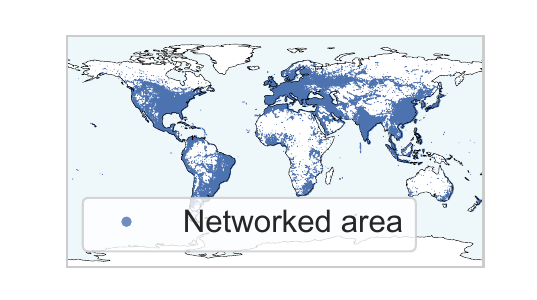}
        \captionsetup{width=0.95\linewidth}
        \caption{Networked regions based on human settlement theory.}
        \label{fig:global_networked}
    \end{minipage}
    \hfill
    \begin{minipage}[t]{0.23\textwidth}
        \centering
        \includegraphics[width=\textwidth]{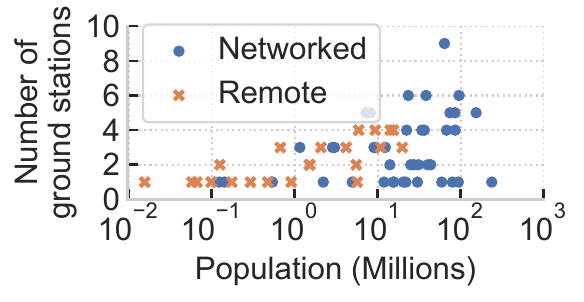}
        \captionsetup{width=0.95\linewidth}
        \caption{Number of Starlink ground stations per population group.}
        \label{fig:GS_per_population}
    \end{minipage}
\end{figure}

\textbf{GS deployment-demand mismatch.} We notice that the ineffectiveness of ISL-based algorithms is caused by a core contradiction in the current LSN structure. \textit{The primary objective of serving remote regions in LSNs conflicts with the practical constraints of GS deployment}. The distribution of GSs and population density is illustrated in Figure~\ref{fig:population_and_gs}, using Starlink, the leading LSN provider, as an example. It is evident that the deployment of GSs is highly biased and related to urbanization. Furthermore, Figure~\ref{fig:global_networked} illustrates the global distribution of networked areas, estimated using human settlement theory, which characterizes modernization by the relationship between population density and nighttime light intensity~\cite{sutton2001census}. Combined with Figure~\ref{fig:population_and_gs}, this reveals that \textit{large populations without Internet access, who are the primary potential users of LSNs, still lack sufficient GS coverage}. Specifically, we partition the Earth's surface into $1{,}500$~km$^2$ grids, approximately matching the coverage area of a GS~\cite{mohan2024multifaceted}. As shown in Figure~\ref{fig:GS_per_population}, many remote population clusters of about $10$ million people are served by only one or two GSs, whereas other networked regions enjoy substantially richer GS availability, with access to more than six GSs.

\textbf{Infrastructure constraints and design goals.} This conflict stems from two factors: (i) Socio-economic and regulatory constraints. Deploying GSs in remote regions faces high deployment costs, limited local infrastructure, spectrum-licensing hurdles, and geopolitical constraints. As a result, underdeveloped regions such as Africa remain in the early stages of GS deployment, with only two Starlink GSs recently activated in Nigeria\footnote{\url{https://dishycentral.com/starlink-ground-station-locations}}. (ii) Data center proximity constraint. Modern 400ZR-based data center interconnects typically span 80-120~km~\cite{packetlight400zr}, limited by optical impairments, signal-to-noise ratio, and cost. Consequently, LSN infrastructures, particularly GSs, are usually colocated with data centers to ensure low-latency, stable communication. However, because data centers are concentrated in developed regions, GS coverage is saturated in North America, eastern South America, Europe, and Australia, as shown in Figure~\ref{fig:population_and_gs}.

Under this skewed GS deployment, the nearest selection approach overloads the already sparse GS resources in remote regions. Although the nearest-available strategy can partially alleviate this by redirecting traffic to more distant GSs, the uneven GS distribution often forces excessively long rerouting paths, leading to unsustainable onboard energy consumption. Thus, we derive our observation as follows.

\refstepcounter{observation}
\textbf{Observation~\theobservation}\label{ob_1}: The ISL-based algorithm cannot resolve congestion in GS resource-scarce regions and suffers from prolonged ISL rerouting paths due to biased GS deployment.

\subsection{Lessons from PCO Delivery}

\textbf{Performance of PCO delivery.} Similarly, we evaluate PCO delivery combined with two GS selection methods. As expected, PCO delivery consumes significantly less energy than ISL-based routing, with an average of $6.73\times$ lower energy consumption. However, we observe that PCO delivery often fails because satellites in traffic-intensive regions have their onboard 1~TB recorders~\cite{nisarSSR2025} fully filled or their batteries completely depleted, while satellites in neighboring orbits remain idle. Specifically, at traffic intensity level $5$, PCO exhibits a failure rate of $48.71\%$ with the nearest-available approach and an even higher rate of $63.03\%$ with the nearest-selection approach due to concentrated traffic allocation. Such a high failure rate renders PCO delivery theoretically attractive but practically unacceptable.

\begin{figure}[t]
     \centering
     \includegraphics[width=\linewidth]{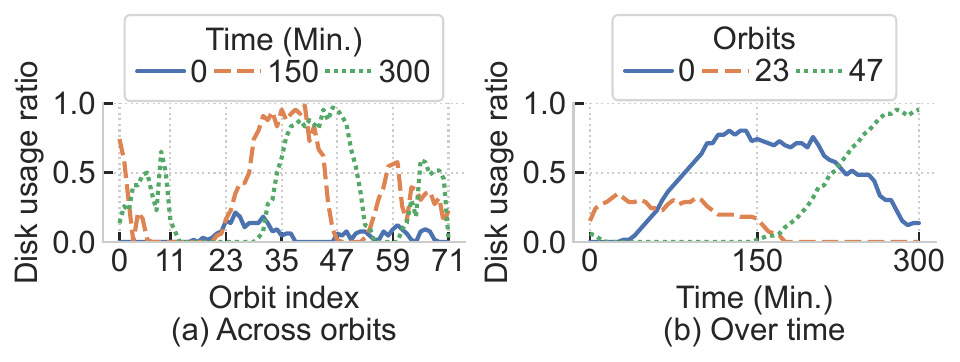}
     \caption{Satellite onboard disk usage ratios across orbits and time.}
     \label{fig:moti_umbra_orbit_cong}
\end{figure}

\textbf{Inefficient onboard resource utilization.} After analyzing satellites' onboard resource usage, we observe that \textit{naive PCO delivery inevitably congests the orbits above target GSs, while neighboring orbits remain mostly idle}. Since PCO requires satellites to pass within the elevation range of target GSs to offload data, the orbits directly overhead become highly congested as the \textit{necessary paths} to the destinations. Figure~\ref{fig:moti_umbra_orbit_cong}(a) shows the average disk usage ratio per orbit at 0, 150, and 300 minutes. At 0 min, most orbits have low storage ratios since satellites initially carry no data. By 150 minutes, satellites in orbit indices 23 to 47 become heavily congested, with average disk usage ratios reaching 0.98, while those in other orbits (2 to 22) remain entirely idle. This pattern occurs when orbits 23 to 47 pass over GS-dense regions, while the others mainly cover oceans.

The congested orbits are also dynamic in the temporal domain due to the Earth's rotation, which makes satellites appear to move westward from the GS perspective, as shown in Figure~\ref{fig:stable_orbit}. As a result, the \textit{necessary paths} to target GSs shift to neighboring orbits, causing peak traffic to gradually move toward higher orbit indices between 150 and 300 minutes, as shown in Figure~\ref{fig:moti_umbra_orbit_cong}(a). This pattern also appears within individual orbits over time, as shown in Figure~\ref{fig:moti_umbra_orbit_cong}(b): three example orbits become congested when passing GS-dense regions and return to idle when traveling above oceans and remote areas. Therefore, the lesson we derive from PCO delivery is as follows.

\refstepcounter{observation}
\textbf{Observation~\theobservation}\label{ob_2}: The existing methods overlook the orbital congestion issue inherent to PCO delivery, leaving resources underutilized in both the spatial and temporal domains.

\subsection{Lessons from Spatiotemporal Dynamics in Space-Ground Networks}

\begin{figure}[t]
     \centering
     \includegraphics[width=\linewidth]{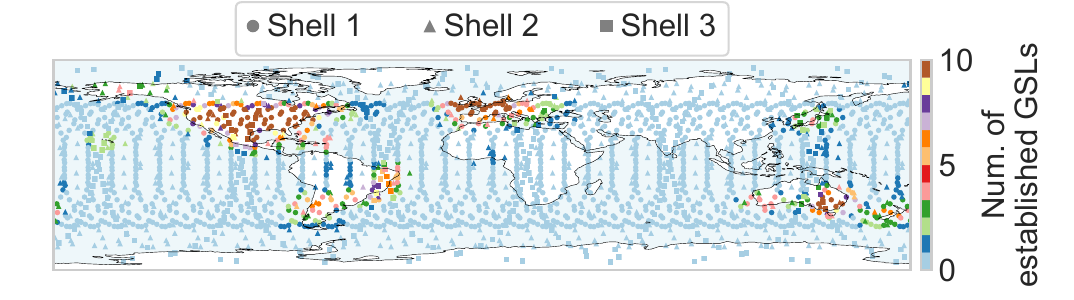}    
     \caption{Satellite distribution and number of established GSLs.}
     \label{fig:sats_visible_gss}
\end{figure}

\begin{figure}[t]
     \centering
     \includegraphics[width=\linewidth]{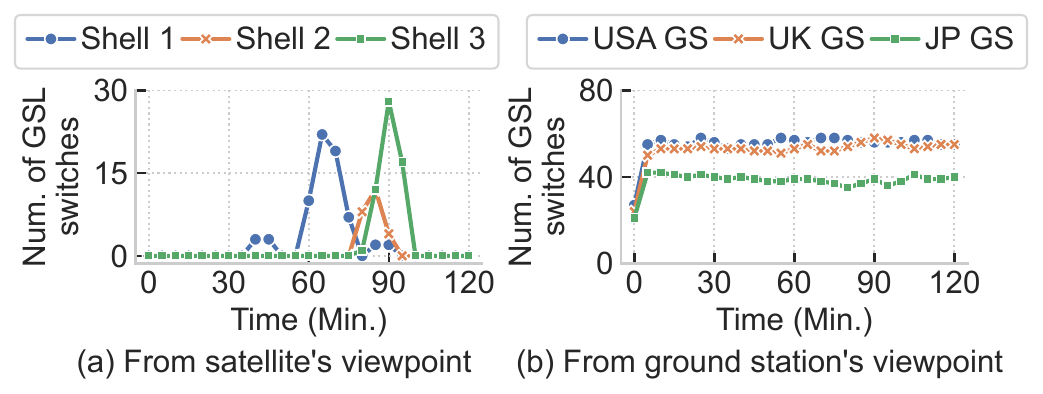}
     \caption{Number of GSL switches from satellite and ground station viewpoints.}
     \label{fig:sats_gss_visibility}
\end{figure}

During the trace-driven study, we identified another potential challenge in control plane implementation. \textit{LSNs involve massive numbers of edge connectivity and node position updates}, which substantially complicate both problem modeling and the solution design.

\textbf{Spatiotemporal dynamics of GSL links.} As shown in Figure~\ref{fig:sats_visible_gss}, a satellite snapshot with a colormap reports the number of established GSLs for each satellite, demonstrating that the spatial distribution of GSLs is highly dynamic. GSL dynamics also manifest in the temporal domain. During an orbital period, a satellite experiences frequent GSL switches while passing GS-dense regions, as shown in Figure~\ref{fig:sats_gss_visibility}(a) for three satellites in different Starlink shells. From the GS perspective, GSL switches are similarly frequent and often sustained, driven by the uniform distribution of the Walker constellation\footnote{\url{https://en.wikipedia.org/wiki/Satellite_constellation\#Walker_Constellation}}, as illustrated in Figure~\ref{fig:sats_gss_visibility}(b) using GSs in the USA, the UK, and Japan.

\begin{figure}[t]
    \centering
    \begin{minipage}[t]{0.23\textwidth}
        \centering
        \captionsetup{width=0.95\linewidth}
        \includegraphics[width=\textwidth]{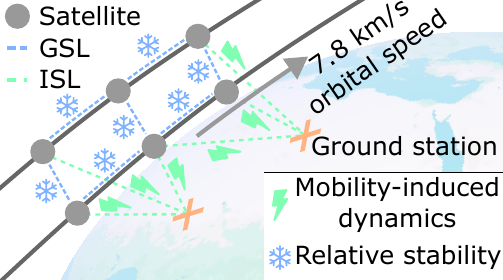}
        \caption{Dynamic GSL and stable intra-orbit ISL connectivity.}
        \label{fig:dynamic_satellite}
    \end{minipage}
    \begin{minipage}[t]{0.23\textwidth}
        \centering
        \captionsetup{width=0.95\linewidth}
        \includegraphics[width=\textwidth]{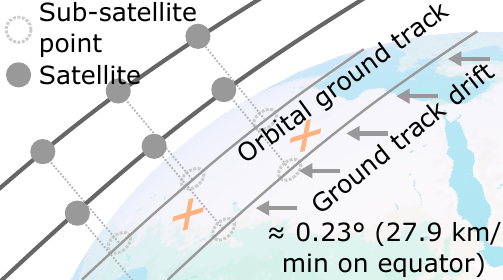}
        \caption{Shift in orbital ground tracks caused by Earth's rotation.}
        \label{fig:stable_orbit}
    \end{minipage}
    \hfill
\end{figure}

Due to the identical altitude, eccentricity, and inclination in a Walker constellation, satellites in the same orbit follow \textit{similar trajectories} and maintain fixed relative positions, a property that also holds for adjacent satellites in neighboring orbits over short time periods. This results in \textit{stable intra-orbit ISL connections} and relatively stable inter-orbit ISL connections between neighboring orbits, as shown in Figure~\ref{fig:dynamic_satellite}. Similarly, the set of orbits visible from a GS remains relatively stable over a time interval, since the Earth's rotation introduces only about $27.9$~km of displacement per minute at the equator, as shown in Figure~\ref{fig:stable_orbit}. Given that a GS can extend coverage to a $2{,}500$~km diameter circle~\cite{mohan2024multifaceted}, the visibility window of an orbit can extend up to $90$ minutes. Therefore, we can conclude the following observation.

\refstepcounter{observation}
\textbf{Observation~\theobservation}\label{ob_3}: Both ISL and GSL connectivity become more stable from an orbital viewpoint.

\subsection{Takeaways}

From the above experiments and analyses, we draw the following insights. (i) An orbital viewpoint modeling framework is required to capture space-ground dynamics while controlling modeling complexity. (ii) A novel GS selection algorithm is needed to address the biased GS deployment, while avoiding prolonged ISL rerouting. (iii) A routing algorithm is necessary to mitigate orbital congestion in PCO delivery while fully utilizing orbital resources. Motivated by these insights, we first formulate the required models in \S\ref{sec:modeling} and subsequently present OrbitTransit in \S\ref{sec:method}.

%% file: sections/3-formulation.tex
\section{LSN Topology Modeling}
\label{sec:modeling}

\subsection{LSN Topology and Task Model}

\textbf{Constellation.} We denote the LSN constellation as $(V, E)$, where $V = \{sat_k \mid k = 1 \dots N_k \}$ is the set of satellites and $E$ is the set of edges representing ISL and GSL links. For ISL establishment, we follow the standard strategy~\cite{deng2021ultra, chen2022load} in which each satellite maintains $4$ ISL connections with its neighbors, two within the same orbit and two with adjacent orbits. We then define the $G = \{g_j \mid j = 1 \dots N_j\}$ as the set of all ground stations. For GSL establishment, a ground station $g_j$ forms a GSL link with a satellite if the elevation angle is greater than $25\degree$~\cite{mohan2024multifaceted}.

\textbf{LSN delivery task.} We define the time series in our model as $T = \{1, \dots, t, \dots, N_t\}$. The delivery tasks in the LSNs initialized at time $t$ can be expressed as $S(t) = \{s_i^t \mid i = 1 \dots N_i\}$, where each task $s_i^t$ is represented as follows:

\begin{equation}
\label{eq:task_define}
s_i^t = (sat, d, \tau)
\end{equation}
$sat$ represents the source satellite when sensing data, or the first-hop satellite of a task when it originates from UE. $d$ denotes the data volume, and $\tau$ indicates the task deadline. 

Each task needs to be assigned to a target GS for data offloading. We use a binary variable to represent the responsibility of GSs for receiving tasks, defined as follows:
\begin{equation}
x^{j}(s_i^t) = \{0, 1\}
\end{equation}
The parameter $x^{j}(s_i^t)$ is $1$ if the $s_i^t$ task is assigned to the $j$-th ground station; otherwise, it is $0$.

Similarly, to formulate each satellite’s responsibility for tasks $s_i^t$, we introduce two binary variables to represent task responsibility for each satellite in $V$, formulated as follows:
\begin{equation}
y_{\hat{t}}^k(s_i^t) = \{0, 1\}, z_{\hat{t}}^k(s_i^t) = \{0, 1\}, \forall \hat{t} \in[t, N_t]
\end{equation}
The variable $y_{\hat{t}}^k(s_i^t)$ is set to 1 if $sat_k$ is involved in the transmission of the $s_i^t$ task at time $\hat{t}$. Similarly, the variable $z_{\hat{t}}^k(s_i^t)$ is set to 1 if $sat_k$ stores the transmission data of the $s_i^t$ task at time $\hat{t}$, and 0 otherwise. Here, $\hat{t}$ belongs to $[t, T]$ because task $s_i^t$ can be processed only after its initialization.

\subsection{Energy Model}

\textbf{Life consumption model.} Due to constrained energy budgets and the high cost of satellite battery replacement, onboard power must be used intelligently. The depth of discharge (DoD) is adopted to evaluate battery life consumption and is defined as follows:
\begin{equation}
DoD_k^{t} = \frac{B_{max} - B_k^{t}}{B_{max}},
\end{equation}
\begin{equation}
B_k^{t} = max(0, E_0^{k, t} + E_+^{k, t} - E_{trans}^{k, t} - E_{IO}^{k, t})
\end{equation}
$B_{\text{max}}$ denotes the maximum onboard battery capacity, and $B_k^{t}$ represents the current battery level of satellite $sat_k$ at time $t$. $E_0^{k, t}$ and $E_+^{k, t}$ represent the current energy level and the energy harvested by the solar panel during time $t$, respectively. The terms $E_{trans}^{k, t}$ and $E_{IO}^{k, t}$ denote the energy consumed for data transmission and storage I/O operations, respectively. They can be formulated as follows:

\begin{equation}
E_{trans}^{k, t} = \sum_{s_i^t \in S(t)} y_{\hat{t}}^k(s_i^t) \cdot d  \cdot \kappa, 
\end{equation}
\begin{equation}
E_{IO}^{k, t} = \sum_{s_i^t \in S(t)} z_{\hat{t}}^k(s_i^t) \cdot d \cdot \zeta
\end{equation}

The two equations compute the total energy consumption at time $t$ by summing all task responsibilities and multiplying them by the energy factors: $\kappa$ Wh/GB for transmission and $\zeta$ Wh/GB for I/O operations, respectively.

In practice, satellites should not deplete all of their energy in order to maintain operational availability and functionality. A minimum battery level is defined to prevent such situations from occurring:
\begin{equation}
\label{eq:minimal_dod_constraint}
\frac{B_k^{t}}{B_{max}} \ge \mathbb{D}, \forall sat_k \in V, t \in T
\end{equation}

The life consumption of a battery is calculated based on the DoD difference between charges and discharges~\cite{yang2016towards}, and it increases exponentially as the difference grows:

\begin{equation}
\label{eq:life_consump}
\mathcal{L}_k^t = e^{max(0, DoD_k^t - DoD_k^{t-1})} - \mathbbm{1}_{DoD_k^t \leq DoD_k^{t-1}}, 
\end{equation}
The symbol $\mathbbm{1}_x$ evaluates to $1$ when the condition $x$ is satisfied and $0$ otherwise. It is used to set the life consumption to $0$ when the previous DoD is greater than the current DoD, indicating a charging event.

\subsection{Constraints and Problem Formulation}

\textbf{GS capacity.} Since each GS is equipped with multiple phased-array antennas, and each antenna can communicate with multiple satellites through time multiplexing~\cite{mohan2024multifaceted, li2024stable}, we focus on the total volume of traffic handled by the GS rather than the number of transmission tasks:

\begin{equation}
\label{eq:gs_cap}
\sum_{s_i^t \in S(t)} x^{j}(s_i^t) \cdot d \le Cap_{g_j}, \\
\forall\; t \in T,\; g_j \in G.
\end{equation}

The summation accumulates the traffic volumes of all tasks that have been assigned to ground station $g_j$ at time $t$. In summary, each GS should receive a total transmission volume that remains within its capacity $Cap_{g_j}$ during any time interval $t$ to $t+1$.

\textbf{Satellite capacity.} Each satellite must not transmit data exceeding its throughput capacity, and its carried data must remain within the onboard storage limits:  

\begin{equation}
\label{eq:sat_cap_cons}
\sum_{s_i^t \in S(t)} y_{\hat{t}}^k(s_i^t) \cdot d \le Cap_{sat_k}, \forall sat_k \in V, t \in T
\end{equation}
\begin{equation}
\label{eq:sat_store_cons}
\sum_{s_i^t \in S(t)} z_{\hat{t}}^k(s_i^t) \cdot d \le Store_{sat_k}, \forall sat_k \in V, t \in T
\end{equation}

\textbf{Task requirements.} All the tasks within $S$ needed to be assigned to a valid GS and have been processed, and their deadline should be guaranteed:

\begin{equation}
\label{eq:task_finish_cons}
\sum_{j \in G} x^{j}(s_i^t) = 1, \forall s_i^t \in S(t), \forall t \in T
\end{equation}
\begin{equation}
\label{eq:task_deadline_cons}
\mathcal{D}_{s_i^t} \le \tau, \forall s_i^t, \forall t \in T
\end{equation}
The first equation states that each task $s_i^t$ needs to be assigned to a valid GS. $\mathcal{D}_{s_i^t}$ denotes the delivery time of task $s_i^t$, which should less than the deadline. The $\mathcal{D}_{s_i^t}$ will be formulated later using the orbit-as-node model.

Given these constraints, our goal is to determine the appropriate GS for each task, as well as the routing path and delivery method (e.g., $x^{j}(s_i^t)$, $y_{\hat{t}}^k(s_i^t)$, and $z_{\hat{t}}^k(s_i^t)$), in order to minimize the overall battery life consumption of the LSN and the average task delivery time:

\begin{equation}
\begin{gathered}
\label{eq:opt_target}
\textbf{minimize} \sum_{sat_{k} \in V, t \in T} \mathcal{L}_k^t + \sum_{s_i^t \in S(t), t \in T} \mathcal{D}_{s_{i}^t}, \\
s.t. \quad (\ref{eq:task_define})-(\ref{eq:task_deadline_cons}),
\end{gathered}
\end{equation}

%% file: sections/4-method.tex
\section{OrbitTransit Methodology}
\label{sec:method}

\begin{figure}[t]
     \centering
     \includegraphics[width=\linewidth]{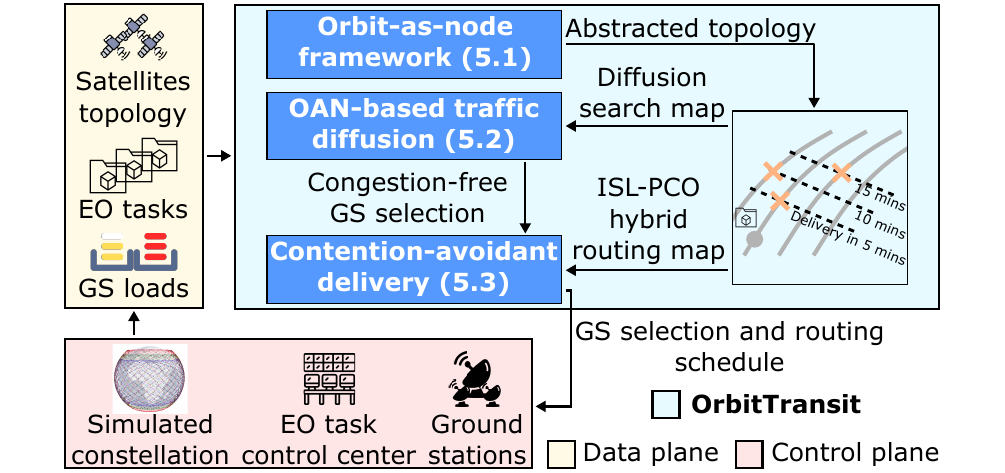}
     \caption{The overview of the OrbitTransit system.}
     \label{fig:framework_overview}
\end{figure}

\begin{figure}[t]
     \centering
     \includegraphics[width=\linewidth]{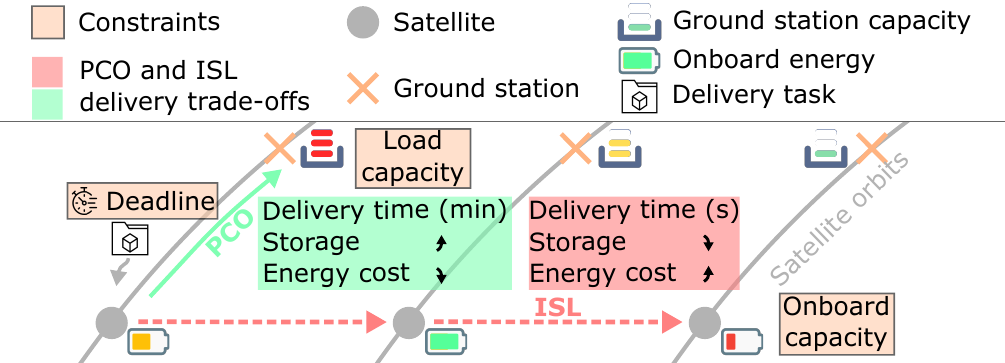}
     \caption{Constraints and trade-offs in ISL-PCO hybrid space delivery problem.}
     \label{fig:algo_tradeoffs}
\end{figure}

In this section, we introduce OrbitTransit, which consists of three key components motivated by the preceding analysis: Orbit-as-Node (OAN) modeling, OAN-based traffic diffusion, and contention-avoidant delivery.

\textbf{Framework overview.} The overview of OrbitTransit is shown in Figure~\ref{fig:framework_overview}. Based on the constellation configuration, the OAN framework abstracts the basic modeling unit from satellites to orbits, providing a simplified orbit-level search space and delivery-time formulation for the following components. The traffic diffusion component then determines the optimal offloading GS for each EO task collected from the EO task control center. Finally, the contention-avoidant delivery module selects the optimal routing path for each EO task and returns the delivery scheme to the GS for execution.

\textbf{Optimization goals.} As shown in Figure~\ref{fig:algo_tradeoffs}, a balanced GS load distribution needs to be achieved to overcome the biased GS deployment mentioned in Observation~\ref{ob_1}. Moreover, the advantages of ISL-based and PCO routing in terms of delivery time, onboard storage, and energy consumption need to be jointly exploited and optimized to address the limitations mentioned in Observation~\ref{ob_1} and Observation~\ref{ob_2}.

\begin{figure}[t]
     \centering
     \captionsetup{belowskip=-5mm, aboveskip=0mm,}
     \includegraphics[width=\linewidth]{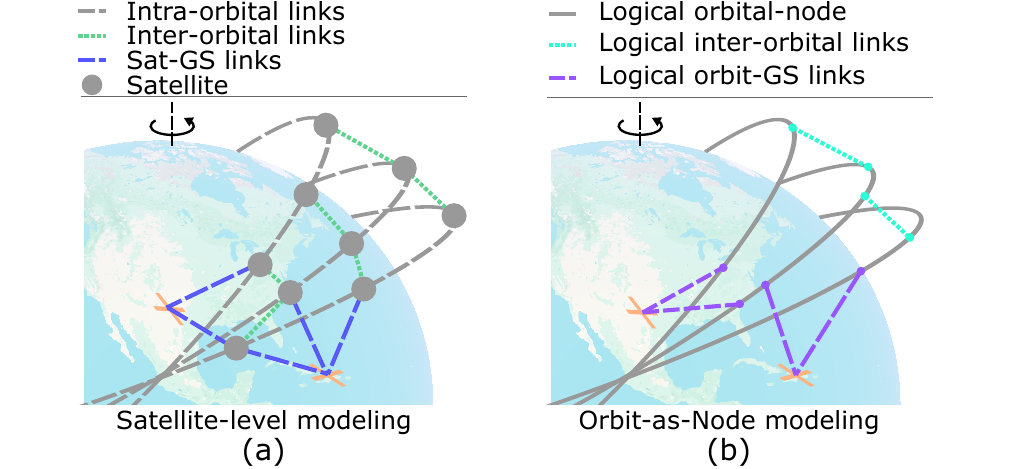}
     \caption{(a) Satellite-level modeling; (b) Orbit-as-node modeling.}
     \label{fig:orbit_as_node}
\end{figure}

\subsection{Orbit-as-Node Framework}
\textbf{OAN overview.} As shown in Figure~\ref{fig:orbit_as_node}(a), satellite-level modeling introduces massive node and edge updates in LSNs. Based on Observation~\ref{ob_3}, we simplify the model, as illustrated in Figure~\ref{fig:orbit_as_node}(b). In this abstraction, each logical orbital node maintains a single GSL to its connectable GSs and a single inter-orbit ISL to its neighboring orbits. These logical links collapse the massive GSL and ISL sets \textit{without affecting graph connectivity}, based on two principles: (i) two neighboring orbits are reachable through inter-orbit links~\cite{mclaughlin2023grid, wang2007topological}; (ii) if one satellite in an orbit is connected to a GS, other satellites in the same orbit can reach this GS via intra-orbit ISLs.

\textbf{Delivery path in OAN model.} We operate at the orbit level rather than per-satellite destinations, \textit{since any satellite on an orbit can reach any point on that orbit via ISL or PCO delivery}. This abstraction greatly reduces routing complexity: with the first Starlink shell (22 satellites per orbit), the routing table shrinks by a factor of $22\times22$ when routing orbit to orbit instead of satellite to satellite. Moreover, the OAN routing \textit{does not compromise path optimality}: (i) the number of inter-orbit hops to the destination orbit is unchanged; (ii) once on the destination orbit, the number of intra-orbit hops to the destination satellite is also unchanged. These properties follow from the grid-like topology of LSNs, where the total path length is invariant to whether the path first traverses along an orbit or across orbits.



\textbf{OAN model formulation.} We define $O = \{o_l \mid l = 1 \dots N_l\}$ as the set of orbits arranged in sequential order. For example, the adjacent orbits of $o_1$ are $o_{N_l}$ and $o_2$. The sets $V(o_l)$ and $G(o_l)$ represent the satellites within orbit $o_l$ and the ground stations covered by this orbit, respectively. As mentioned earlier (Figure~\ref{fig:dynamic_satellite} and Figure~\ref{fig:stable_orbit}), these two sets are relatively stable, with their ground tracks influenced only by the Earth's rotation. 

We then formulate the PCO delivery time as $\mathcal{O}_{sat_k}^{t}(g_j)$, which represents the orbital flight time required for $sat_k$ at time $t$ to reach the elevation range of ground station $g_j$, where a GSL can be established. This delivery time is calculated based on the orbital speed and the angular distance to the target elevation range.

\begin{figure}[t]
    \centering
    \begin{minipage}[t]{0.23\textwidth}
        \centering
        \includegraphics[width=\textwidth]{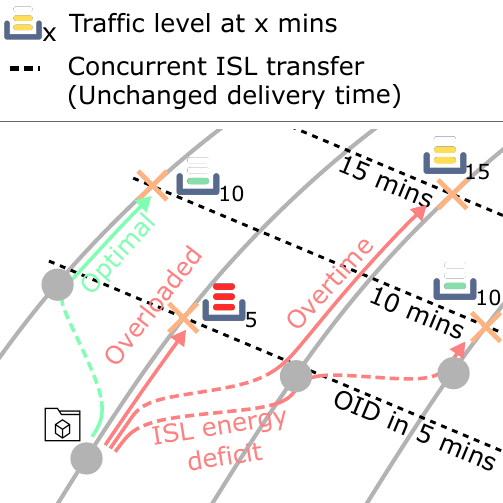}
        \captionsetup{width=0.95\linewidth}
        \caption{A traffic diffusion instance.}
        \label{fig:dist_visual}
    \end{minipage}
    \begin{minipage}[t]{0.23\textwidth}
        \centering
        \includegraphics[width=\textwidth]{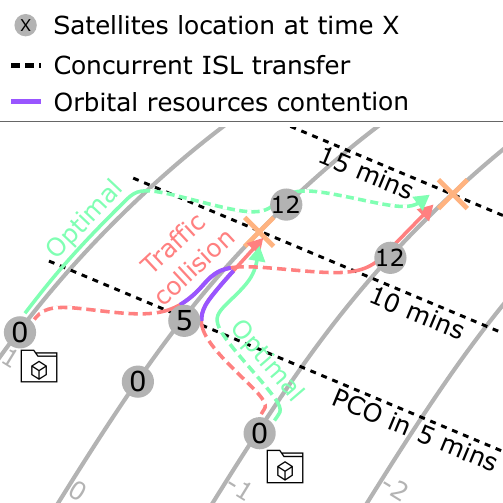}
        \captionsetup{width=0.95\linewidth}
        \caption{An orbital contention instance.}
        \label{fig:contention_visual}
    \end{minipage}
    \hfill
\end{figure}

\textbf{Delivery time formulation.} Then the task delivery time $\mathcal{D}_{s_i^t}$ can be formulated using the PCO delivery time and the binary variables we defined above:
\begin{equation}
\label{eq:task_delivery_time}
\mathcal{D}_{s_i^t} = t + \sum_{t \in T}\sum_{sat_k \in V} z_{\hat{t}}^k(s_i^t) \sum_{g_j\in G} x^{j}(s_i^t) \mathcal{O}_{sat_k}^{t}(g_j)
\end{equation}

The first two summations identify all satellites $sat_k$ assigned to store and perform PCO delivery of task $s_i^t$ across all time. The third summation locates the ground station $g_j$ to which task $s_i^t$ is offloaded, extracts the corresponding PCO delivery time, and aggregates it accordingly. In summary, this equation selects all satellites involved in the PCO delivery of task $s_i^t$ and sums their corresponding delivery times, since the delivery may be completed by multiple satellites.

\textbf{Concurrent ISL-PCO transmission}. Since a satellite maintains relatively stable inter-orbit ISLs with its neighbors while traversing its orbit, \textit{PCO delivery and ISL transmission can occur concurrently}, as illustrated in Figure~\ref{fig:dist_visual}. Taking the optimal light-green path as an example, the data crosses from orbit 0 to orbit 1 via an ISL while simultaneously approaching the target GS on orbit 1 through PCO delivery. Therefore, the ISL transmission time is inherently absorbed into our model, as the PCO delivery time is dominant.

\begin{algorithm}[t]
\caption{OAN-based traffic diffusion} \label{alg:mission_scheduling}
\label{al:traffic_diffusion}
\begin{algorithmic}[1]
\Statex \textbf{Input:} $T, (V, E), G, S(t)$
\Statex \textbf{Output:} $\{x^{j}(s_i^t) | \forall s_i^t \in S(t), t \in T,  g_j \in G\}$
\State $S_{assign} = \{\}$
\For {$t \in T$}
    \For {$\delta \in \{0, 1, -1, 2, -2 ,\dots\}$} \Comment{Traffic diffusion} \label{algline:traffic_diffusion}
        \For {$s_i^t \in S(t)$}  
        \State $sat, d, \tau \leftarrow s_i^t$ \label{algline:task_initial}
        \State $o_l(sat) := \text{the unique } o \in O \text{ with } sat \in V(o)$
        \State $\tilde{o} := o_{l+\delta}$ \label{algline:tar_orbit}
            \For {$g_j \in G(\tilde{o})$} \label{algline:iterate_gs} \Comment{Ascending by distance to $sat$}
                \State  $m_{g_j} := \sum_{s_i^t \in S_{assign}} x_j(s_i^t)\cdot d$ \Comment{Load of GS $g_j$}
                \If {$\mathcal{O}_{sat}^t(g_j) \le \tau \;\land\; m_{g_j} + d \leq Cap_{g_j}$} \label{algline:gs_condition_check}
                    \State $x^{j}(s_i^t) = 1; S_{assign} := S_{assign} \cup \{s_i^t\}$ \label{algline:add_assign}
                    \State $S(t) := S(t) \setminus \{s_i^{t}\}$ \label{algline:remove_complete}
                \EndIf
            \EndFor
        \EndFor
        \If {$S(t) = \emptyset$} \Comment{All tasks assigned}
            \State Break 
        \EndIf
    \EndFor
\EndFor
\end{algorithmic}
\end{algorithm} 

\subsection{OAN-based Traffic Diffusion} With the defined OAN model, we can determine the offload GS $g_j$ for each task $s_i^t$ (represented by the binary variable $x^j(s_i^t)$), as shown in Algorithm~\ref{alg:mission_scheduling}. 

\textbf{Task-GS assignment}. For each time $t$ and every task $s_i^t$ initialized at that time, we identify the orbit $o_l$ to which the source satellite $sat$ belongs (Lines~\ref{algline:task_initial}-\ref{algline:tar_orbit}) using the OAN model. The variable $\tilde{o}$ denotes the orbit currently being searched for suitable GSs. Line~\ref{algline:iterate_gs} evaluates each GS in ascending order of distance to $sat$ to \textit{ensure that tasks are completed as early as possible}. Line~\ref{algline:gs_condition_check} verifies whether the PCO delivery time is less than the deadline $\tau$ and whether the GS $g_j$ still has sufficient capacity to handle the task, which \textit{enforces the constraints in Eq.~(\ref{eq:gs_cap}) and Eq.~(\ref{eq:task_deadline_cons})}. Any task passing these checks is removed from the set $S(t)$ (Lines \ref{algline:add_assign}-\ref{algline:remove_complete}).

\textbf{Orbit-wise traffic diffusion}. If a task cannot be assigned because all GSs in orbit $\tilde{o}$ are either congested or unable to meet the task deadline, the algorithm diffuses the traffic to neighboring orbits through ISLs (Line~\ref{algline:traffic_diffusion}), as illustrated in Figure~\ref{fig:dist_visual}. Starting from orbit 0, it sequentially evaluates orbit $1$, orbit $-1$, orbit $2$, and orbit $-2$ until an optimal GS is found, ensuring that the \textit{delivery path traverses the minimal number of ISLs between orbits}. This procedure is guaranteed to terminate as long as the aggregate GS capacity exceeds the total data volume to be transmitted, thereby ensuring the constraint in Eq.~(\ref{eq:task_finish_cons}). This mechanism is also tailored to address the orbital congestion issue discussed in Observation~\ref{ob_2}.

\subsection{Contention-Avoidant Delivery}
Given the congestion-free GS selection, the contention-avoidant delivery determines the optimal PCO-ISL routing path as shown in Algorithm~\ref{alg:transfer_scheduler}. For each task $s_i^t$, we first extract the relevant information and the candidate GS chosen by traffic diffusion algorithm (Lines \ref{alg2line:task_initial}-\ref{alg2line:candi_gs}).

\textbf{Support for real-time tasks.} We then check whether the delivery time $\mathcal{D}_{s_i^t}$ allows the task to reach the selected GS before its deadline. In rare cases, some EO tasks may have deadlines close to real-time~\cite{nasa_earthdata_2024}, in which any PCO-involved delivery cannot satisfy the requirement. For such tasks, we set the delivery mode to ISL-only to guarantee timely completion (Line~\ref{alg2line:isl_only}) and use the space routing algorithm~\cite{yang2016towards, liu2024orbit, chen2022delay, lai2023achieving, pan2025stableroute} to determine the optimal path, thereby ensuring that the OrbitTransit framework is compatible with both real-time and delay-tolerant applications.

\textbf{PCO-only delivery.} If the deadline permits and the target GS $g_j$ lies on the same orbit as $sat$, then $sat$ can deliver the task via PCO without ISLs. The transmission variable $y^{sat}$ is set to $1$ at time $t$ and $\mathcal{D}_{s_i^t}$, and the storage variable $z^{sat}$ is set to $1$ over this interval (Lines~\ref{alg2line:PCO-only_start}–\ref{alg2line:PCO-only_end}). These variables indicate the GSL that picks up and offloads the task data and the satellite that carries the data, respectively.

\textbf{PCO-ISL hybrid delivery.} If the target GS is not covered by the current orbit, the task data must be forwarded to orbit $o_{\delta}$ covering it (Lines~\ref{alg2line:hybrid_start}-\ref{alg2line:set_store_after}). We initialize $t_{isl}$ as $\mathcal{D}_{s_i^t}$ to mark the ISL transmission start time, a critical factor later resolved to avoid the resource contention in Observation~\ref{ob_2}. Before ISL transmission, the source satellite $sat$ carries the data (Line~\ref{alg2line:set_store_before}). Once ISL transmission begins, satellites along path $\pi$ forward the data (Line~\ref{alg2line:set_trasfer_path}). Finally, the last satellite on path $\pi$ carries the data and offloads it to the target GS (Line~\ref{alg2line:set_store_after}), where $tail(\pi)$ denotes the last satellite in the path.

Once all tasks have their $t_{isl}$ initialized, the optimal ISL time $t_{isl}$ is obtained by solving a minimum-cost maximum-flow (MCMF) problem under satellite capacity constraints (Line~\ref{alg2line:MCMF}). This avoids delivery tasks with resource contention, as illustrated in Figure~\ref{fig:contention_visual}. For example, if $t_{isl}=t$ and ISL transmission starts immediately, the satellite in orbit 0 experiences contention between 5 and 10 minutes. This can be resolved if the task in orbit 1 delays its $t_{isl}$ to 10 minutes. The effectiveness is further shown by comparing its optimality gap with the ground truth in the evaluation section.

\begin{algorithm}[t]
\caption{Contention-avoidant delivery} \label{alg:transfer_scheduler}
\label{al:contention_avoid}
\begin{algorithmic}[1]
\Statex \textbf{Input:} $T, (V, E), G, S(t)$
\Statex \textbf{Output:} $\{y_{\hat{t}}^k(s_i^t), z_{\hat{t}}^k(s_i^t) | \forall s_i^t \in S(t), t \in T,  sat_k \in V\}$

\For {$t \in T$}
    \For {$s_i^t \in S(t)$}
        \State $sat, d, \tau \leftarrow s_i^t$ \label{alg2line:task_initial}
        \State $o_l(sat) := \text{the unique } o \in O \text{ with } sat \in V(o)$
        \State $g_j :=$ extract target GS from the $x^{j}(s_i^t) = 1, \forall g_j \in G$\label{alg2line:candi_gs}
        \If {$\mathcal{D}_{s_i^t} \le \tau$}
            \If {$g_j \in G(o_l)$} \label{alg2line:PCO-only_start}
                \State $y_{t}^{sat}(s_i^t) = 1; \, y_{\mathcal{D}_{s_i^t}}^{sat}(s_i^t) = 1$ \Comment{GSL receive/sent} \label{alg2line:gsl_rece_and_sent}
                \State $z_{\tilde{t}}^{sat}(s_i^t) = 1, \forall\tilde{t} \in [t, \mathcal{D}_{s_i^t}] $ \Comment{Data carry} \label{alg2line:PCO-only_end}
            \Else \Comment{Cross orbit data transfer} \label{alg2line:hybrid_start}
                \State $o_{\delta} :=$ the orbit can cover $g_j$
                \State $t_{isl} = \mathcal{D}_{s_i^t}$ \Comment{Time to ISL}
                \State $z_{\tilde{t}}^{sat}(s_i^t) = 1, \forall \tilde{t} \in [t, t_{isl}]$ \label{alg2line:set_store_before}
                \State $\pi :=$ extract ISL path to orbit $o_{\delta}$
                \label{alg2line:trasfer_path}
                \State $y_{t_{isl}}^{\hat{sat}}(s_i^t) = 1, \forall \hat{sat} \in \pi$\label{alg2line:set_trasfer_path}
                \State $z_{\tilde{t}}^{tail(\pi)}(s_i^t) = 1, \forall  \tilde{t} \in [t_{isl}, \mathcal{D}_{s_i^t}]$ \label{alg2line:set_store_after}
            \EndIf 
        \Else   \Comment{ISL-only transmission} \label{alg2line:isl_only}
            \State $\pi :=$ extract ISL path to GS $g_j$
            \State $y_{t}^{\hat{sat}}(s_i^t) = 1, \forall \hat{sat} \in \pi$
        \EndIf
    \EndFor
    \State Solve for $t_{isl}$ for all tasks using the MCMF solver, subject to the constraints in Eqs.~(\ref{eq:sat_cap_cons}) - (\ref{eq:sat_store_cons}). \label{alg2line:MCMF}
\EndFor

\end{algorithmic}
\end{algorithm} 

%% file: sections/5-system_design.tex
\section{System Design}
\label{sec:sys_design}

\textbf{Prototype compatibility.} As shown in Figure~\ref{fig:framework_overview}, OrbitTransit relies on data plane inputs obtainable from existing satellite hardware and ground infrastructure via CCSDS-based telemetry~\cite{Kearney_CCSDS_2014}. Telemetry packets provide satellite status, including state of charge, payload activity, and memory usage, which are periodically collected by the onboard computer and broadcast to GSs. EO tasks are issued by mission control centers (e.g., ESMO~\cite{NASA_ESMO_Webpage}) with target coordinates, observation modes, and timing parameters. Global GS load can also be obtained through distributed ground network management systems~\cite{KSAT_GPD_FinalReport_2025}. Based on these inputs, OrbitTransit generates GS selection and routing schedules, which are forwarded by GSs to the target satellites for execution. To illustrate its compatibility with existing EO workflows, a brief wildfire monitoring example is provided in Appendix~\ref{sec:appendix_wildfire}.

\textbf{Robustness and scalability.} In practice, transmission data and telemetry may be delayed or partially missing due to packet loss or limited bandwidth~\cite{li2024satguard, mohan2024multifaceted, fang2024robust}. For unacknowledged data, OrbitTransit performs opportunistic ISL retransmissions or reroutes traffic when the previous relay moves out of range, by re-executing Lines~\ref{alg2line:hybrid_start}--\ref{alg2line:set_store_after} in Algorithm~\ref{alg:transfer_scheduler}. For delayed telemetry, the Orbit-as-node model is tolerant because scheduling decisions rely on orbital-level aggregates, where errors from any single satellite have limited influence. The aggregated state is updated once delayed telemetry arrives, correcting prior estimates.

The control plane operates as a periodic and event-driven controller, incrementally updating the topology when telemetry indicates significant changes in GS load or orbital resources, or when previously missing telemetry arrives, rather than recomputing routes upon each per-satellite telemetry update. This design bounds control-plane overhead and improves scalability to large-scale LSN constellations.

\textbf{Fault tolerance during PCO delivery.} Due to the unpredictable outer-space environment, a satellite may become unavailable because of solar storms or emergency maneuvers~\cite{liu2024orbit, li2023networking}, preventing it from continuing its data-carrying mission. In such cases, the OrbitTransit reconstructs the remaining path and selects the nearest feasible satellite via intra- or inter-orbital ISLs to resume delivery. If the fallback satellite lacks sufficient storage, the task is temporarily fragmented and forwarded through multiple ISL relays until a valid satellite or target GS becomes available. Similarly, when rerouting leads to deadline violations or an emergency advances the task deadline, delivery is escalated to an ISL-only mode, as shown in Algorithm~\ref{alg:transfer_scheduler}.

\textbf{Adaptation to unpredictable exceptions.} Some commercial GSs, such as Starlink, also act as Internet exchange or peering facilities for terrestrial networks, so their aggregate load is highly dynamic and no longer solely determined by satellite traffic. When the target GS cannot accommodate scheduled offloading due to unexpected external traffic surges, the satellite can temporarily defer offloading until the GS becomes available, since a typical LEO satellite maintains a communication window of about 10 minutes with a GS~\cite{mohan2024multifaceted}. If offloading remains infeasible after this window, the control plane initiates a new mission for OrbitTransit to identify and execute offloading to the next optimal GS.

%% file: sections/6-evaluation.tex
\section{Implementation and Evaluation}
\label{sec:evalution}

\begin{figure*}[t]
     \centering
     \includegraphics[width=\linewidth]{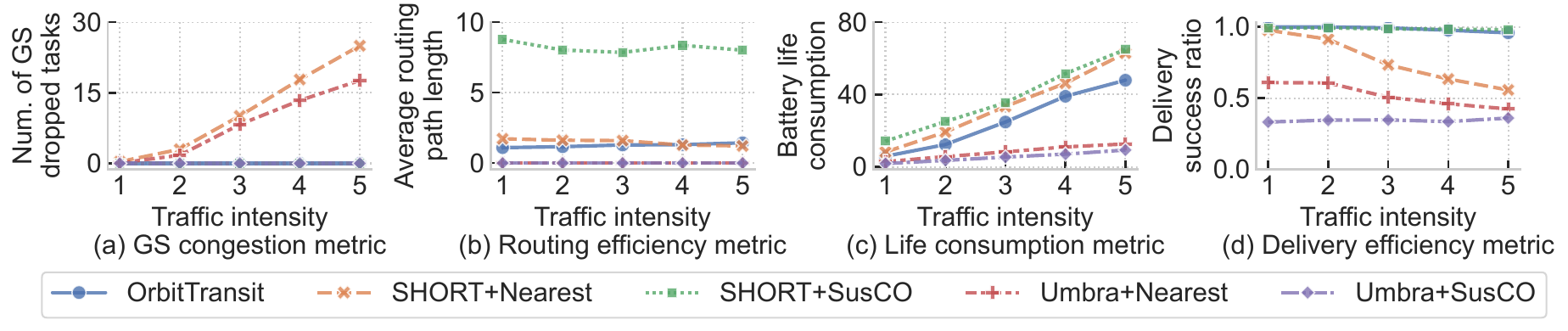}
     \caption{Overall comparison of all baseline combinations across four key evaluation metrics.}
     \label{fig:overall_compare}
\end{figure*}

\subsection{OrbitTransit Implementation}

\textbf{Experiment parameters.} Our satellite constellation model is built from two-line element data~\cite{celestrak2025} for Starlink, OneWeb, and Telesat, with $1\,584$, $720$, and $784$ satellites at altitudes from $550$ to $1{,}200$ km. These form representative constellations with different scales, altitudes, and coverage. Each satellite is equipped with a $5{,}000$~Watt-minute battery~\cite{yang2016towards}, a 1 TB data recorder~\cite{nisarSSR2025}, and solar panels providing $120$~W charging power in sunlight~\cite{liu2024orbit}, with the minimum battery threshold $\mathbb{D}$ set to $0.2$. The energy factors $\kappa$ and $\zeta$ are set to $0.08$~\cite{yang2016towards} and $2.51\times10^{-5}$~\cite{Mercury2021RH3440} Watt-minutes per megabit, representing unit energy consumption for transmission and I/O. Based on~\cite{li2024stable, liu2024orbit, zhai2024seco, cao2023computing}, each satellite has ISL and GSL bandwidth $Cap_{sat}$ of 1~Gbps.

The GS setup is based on infrastructure documented by the FCC~\cite{spacex_services_fcc}, consisting of 165 GSs in total. Following~\cite{mohan2024multifaceted}, a GS establishes a GSL link with a satellite when the elevation angle exceeds $25\degree$. The maximum GS capacity $Cap_{g}$ is 10~Gbps when $8$ phased-array antennas are deployed, as inferred from Starlink’s FCC filings~\cite{FCC2025Starlink40Antennas}. OrbitTransit and the control plane are implemented in Python with 2,500 lines of code based on the PyEphem astronomy library, and all experiments are run on a Linux server with two AMD 7313 CPUs.

\textbf{Delivery tasks.} Based on reports that EO missions can generate 3.9-7~TB of data per day per sensor or satellite~\cite{nesdis_jpss_cloud_2021, nesdis_viirs_2025}, we sample each EO task with a delivery amount $d$ from 2~TB to 10~TB to evaluate baseline performance under varying traffic intensities. EO task deadlines range from minutes to hours depending on urgency. Wildfire detection typically requires delays within 20-30 minutes~\cite{nasa_firms_2022}, whereas long-term monitoring data can tolerate delays up to 3~hours~\cite{nasa_earthdata_2024}. Accordingly, each task is assigned a deadline $\tau$ between 20 and 180 minutes to represent different urgency levels.

\textbf{Evaluation metrics.} We evaluate the baselines from three perspectives: (i) \textbf{GS metrics.} GS load ratio, number of dropped tasks, and queueing delay characterize congestion based on incoming transmission volume and the GS capacity described earlier. (ii) \textbf{Satellite metrics.} Satellite energy consumption, recorder storage utilization, and battery life consumption are estimated using the standard lithium-ion battery model~\cite{yang2016towards}. (iii) \textbf{Task metrics.} Task delivery and failure ratios are analyzed, with failures divided into timeout, GS congestion, and storage overflow. Average routing path length, measured in hops, is also reported to represent routing quality.

\textbf{Comparison baselines.} We select two GS selection methods. (i) \textbf{Nearest}~\cite{ma2023network, tao2023transmitting, zhao2023realtime}: an early-stage Starlink strategy that forwards traffic to the nearest available GS. (ii) \textbf{SusCO}~\cite{chou2025commercial}: a LEO offloading framework that selects low-cost collaborative GS groups to reduce energy consumption and improve capacity. We also include two representative space routing algorithms covering both ISL-based and PCO-like delivery. (i) \textbf{Umbra}~\cite{tao2023transmitting}: a PCO method with a withhold scheme, where satellites retain data when the target GS is congested and later offload it via time-expanded networks. (ii) \textbf{SHORT}~\cite{li2024stable}: a LEO routing algorithm using orbital geodetic addresses to adapt to maneuvers, failures, and dynamic conditions. We evaluate OrbitTransit against all combinations of these GS selection and routing baselines.

\begin{figure*}[t]
     \centering
     \includegraphics[width=\linewidth]{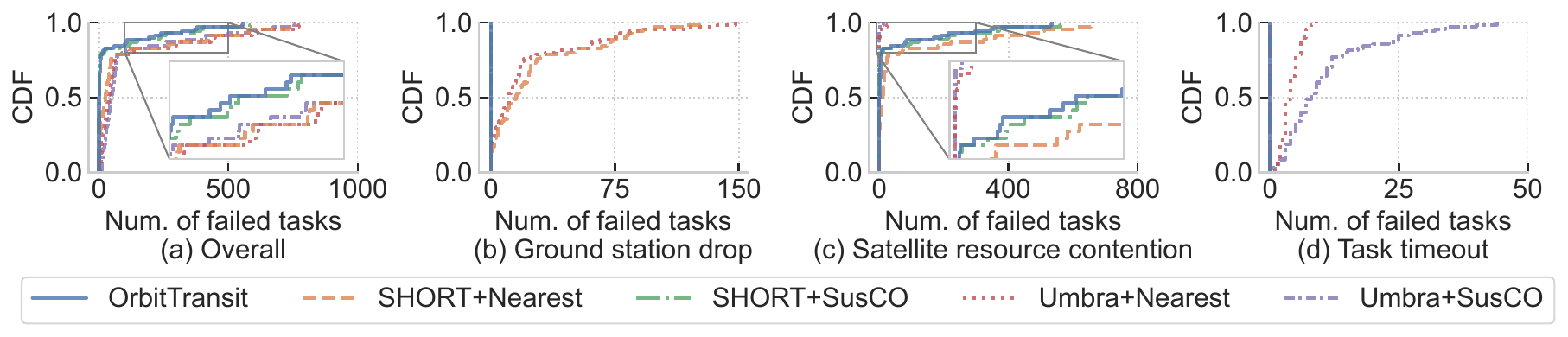}
     \caption{Cumulative distribution of failed delivery tasks categorized by failure reason.}
     \label{fig:evaluation_task_status}
\end{figure*}

\begin{figure}[t]
     \centering
     \includegraphics[width=\linewidth]{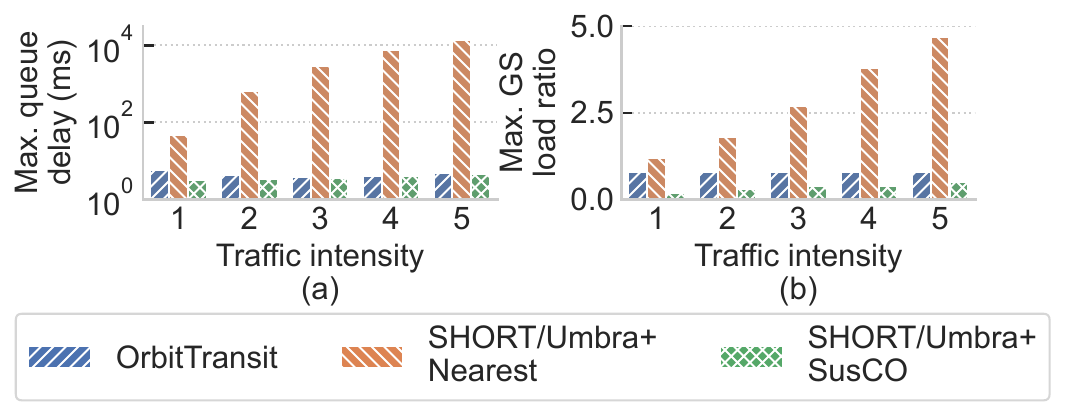}
     \caption{Ground station performance under different selection methods.}
     \label{fig:evaluation_gs_status}
\end{figure}

\subsection{Performance Evaluation} 

\textbf{System-level performance evaluation.} We first evaluate the key metrics in Figure~\ref{fig:overall_compare}. For GS congestion, the nearest GS selection strategy quickly fails due to the imbalance between LSN supply and demand, creating severe bottlenecks at nearby GSs, as shown in Figure~\ref{fig:overall_compare}(a). Although SHORT+SusCO can strategically select suitable GSs to balance the load, the biased GS distribution forces it to choose GSs far from the UE, resulting in a 5.52$\times$ longer ISL propagation path, as shown in Figure~\ref{fig:overall_compare}(b). In contrast, OrbitTransit achieves the best overall performance by alleviating GS congestion through traffic diffusion, while maintaining a routing path length comparable to the ISL-based method SHORT, and only slightly higher than the PCO-like method, Umbra.

The prolonged ISL path of SHORT+SusCO yields the highest battery life consumption, as shown in Figure~\ref{fig:overall_compare}(c). Umbra-based methods minimize battery consumption by avoiding ISL usage, yet \textit{attain only a 43.25\% delivery success ratio} due to frequent onboard resource contention and task timeouts, as shown in Figure~\ref{fig:overall_compare}(d). OrbitTransit and SHORT+SusCO both reach a $100\%$ delivery success ratio at low traffic intensity, but performance declines as traffic approaches constellation and GS capacity limits. Overall, OrbitTransit achieves the highest delivery success ratio while reducing battery consumption by 47.16\% compared to ISL-based routing.

\begin{figure}[t]
     \centering
     \includegraphics[width=\linewidth]{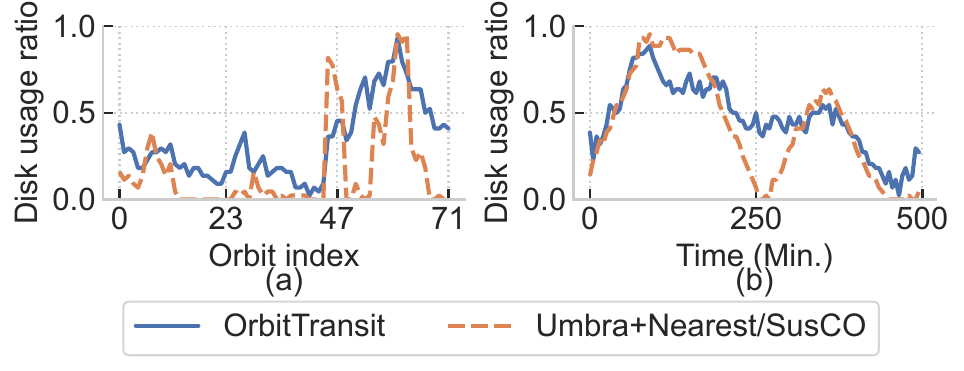}
     \caption{Satellite disk usage ratio under different PCO-like methods.}
     \label{fig:evaluation_sat_orbit_status}
\end{figure}

\begin{figure*}[t]
     \centering
     \includegraphics[width=\linewidth]{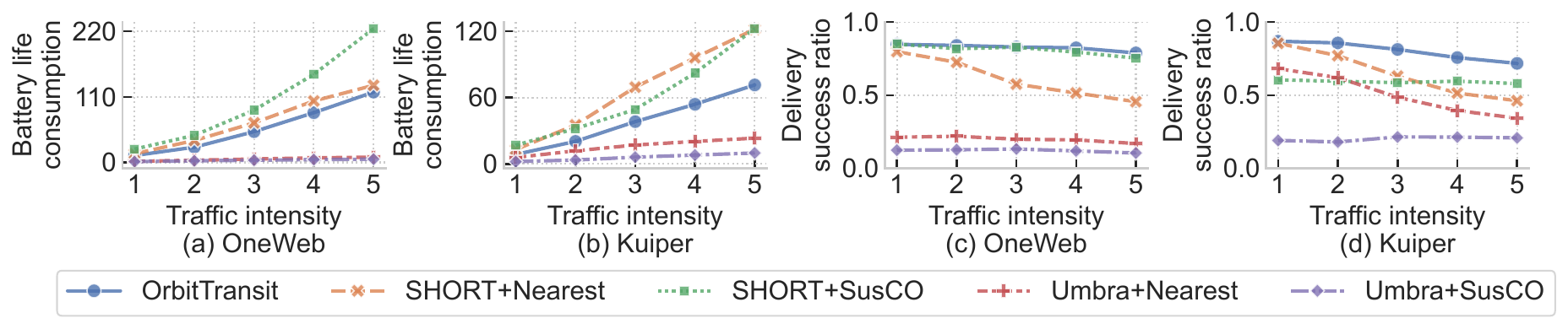}
     \caption{Baseline performance under different constellation configurations.}
     \label{fig:evaluation_lsns_compare}
\end{figure*}

\textbf{Evaluation in task metrics.} Figure~\ref{fig:evaluation_task_status} shows the cumulative distribution of failed tasks under each baseline. OrbitTransit yields $1.09\times$ fewer failed delivery tasks than other baselines (Figure~\ref{fig:evaluation_task_status}(a)), with most failures caused by traffic intensity exceeding constellation capacity. Specifically, all baselines with nearest GS selection consistently suffer from GS congestion, leading to high queue delays and task drops, as shown in Figure~\ref{fig:evaluation_task_status}(b). Moreover, SHORT routing, with intensive ISL usage, often exhausts satellite batteries in hotspot regions, causing resource contention (Figure~\ref{fig:evaluation_task_status}(c)), with $3.34\times$ more failed tasks than other baselines. Finally, as shown in Figure~\ref{fig:evaluation_task_status}(d), Umbra routing frequently misses task deadlines due to the lack of delivery-time management, resulting in delayed and inefficient PCO delivery.

\textbf{Evaluation in GS metrics.} As shown in Figure~\ref{fig:evaluation_gs_status}, the nearest selection strategy quickly degrades as traffic intensity increases because it lacks a traffic distribution mechanism, resulting in a queue delay of $10^4$ ms and a load ratio of 4.7, about $5\times$ the designed capacity. Although SusCO achieves the lowest average queue delay and load ratio by distributing traffic more sparsely, it also leads to longer ISL paths, as shown in Figure~\ref{fig:overall_compare}(b). OrbitTransit maintains a maximum queue delay of 4.75~ms and a load ratio of 80\%, comparable to SusCO in quality of service, but with a shorter path length.

\textbf{Evaluation in satellite metrics.} We compare Umbra with OrbitTransit to evaluate orbital resource utilization. As shown in Figure~\ref{fig:evaluation_sat_orbit_status}(a), high-load orbits are between indices 45 and 68. Umbra uses only a few orbits, leaving neighbors underutilized and causing low overall storage utilization. In contrast, through orbit-wise traffic diffusion (Figure~\ref{fig:dist_visual}), OrbitTransit distributes traffic across neighboring orbits, achieving a higher delivery success ratio (Figure~\ref{fig:overall_compare}(d)) while avoiding overload on individual orbits. In the temporal domain, orbits are effectively utilized only when covering GS-dense regions, causing low utilization for Umbra around 250 minutes, as shown in Figure~\ref{fig:evaluation_sat_orbit_status}(b). OrbitTransit instead maintains higher and more balanced utilization over time via contention-avoidant delivery, scheduling carry-on and offloading to leverage idle satellites and smooth peak traffic.

\textbf{Evaluation in different constellations.} The generality of OrbitTransit across constellations is shown in Figure~\ref{fig:evaluation_lsns_compare}. Battery consumption increases noticeably for OneWeb and Kuiper compared to Starlink, since satellites in smaller constellations must handle more traffic and are more likely to deep discharge. Similarly, all baselines show a steeper decline in delivery success ratio as traffic intensity grows, due to sparser satellite distribution and more frequent onboard resource contention. Overall, OrbitTransit reduces battery consumption by $49.57\%$ and $71.12\%$ on OneWeb and Kuiper compared to SHORT, while improving delivery success ratio by $53.20\%$ and $44.71\%$ compared to all baselines.

\textbf{Evaluation against the optimal solution.} To assess how closely OrbitTransit approaches the ground-truth solution in the GS selection and space routing problems, we employ PuLP~\cite{PuLP_Github} as a linear and mixed-integer programming modeler and use its solution as a benchmark. Due to space constraints, we defer the detailed formulation and experiment results to Appendix \S\ref{sec:appendix_pulp}. The experiment shows that OrbitTransit achieves performance comparable to PuLP while incurring much lower runtime and memory overhead.

\textbf{Evaluation under delayed data plane states.} We also evaluate the robustness of OrbitTransit under delayed data-plane states, as described in Section~\S\ref{sec:sys_design}. The results show that OrbitTransit incurs only a $2.15\%$ degradation in routing efficiency, while remaining comparable in other metrics. Details are provided in Appendix~\S\ref{sec:appendix_delay}.

%% file: sections/7-related_work.tex
\section{Related Work}

Recent research efforts on LSNs span from topology and constellation design~\cite{bhattacherjee2019network, hauri2020internet, singh2021community} to early-stage measurements~\cite{kassem2022browser, michel2022first, tanveer2023making, ma2023network, mohan2024multifaceted, zhao2023realtime}. This trend has further extended to protocol design~\cite{lai2022spacertc, li2021processing, cao2023satcp, vasisht2021l2d2, li2023networking}, onboard task scheduling~\cite{liu2024orbit, chen2021time, shenoy2024cosmac}, and specialized frameworks for applications such as Earth observation~\cite{tao2024known, fu2024fireloc, liu2023multi} and video streaming~\cite{zhao2024low, zhang2024starstream, fang2024robust, zhao2025baroc}.

Due to the mobility and sustainability concerns of satellite platforms, space routing studies have aimed to improve bandwidth, delay, reliability, and battery efficiency. For example, Lai et al.~\cite{liu2024orbit} proposed an ISL-based routing algorithm for battery-aware forwarding, while Chen et al.~\cite{chen2022delay} introduced a cooperative transmission scheme for dynamic topologies and time-varying resources. Meanwhile, prior works~\cite{tao2023transmitting, huang2018envisioned} have explored using satellite mobility for data delivery. In particular, Tao et al.~\cite{tao2023transmitting} proposed a withhold scheduling scheme in which a satellite defers transmission when the visible GS is suboptimal and offloads to a future GS.

Our trace-driven study shows that prior ISL-based routing methods~\cite{yang2016towards, liu2024orbit, chen2022delay, lai2023achieving, pan2025stableroute} cannot fundamentally resolve detoured ISL paths caused by uneven GS deployment and still rely heavily on ISLs. Existing PCO-like delivery studies~\cite{tao2023transmitting, huang2018envisioned} overlook resource contention on congested orbits, leading to inefficient utilization. Motivated by these limitations and our measurements, we propose OrbitTransit, a hybrid PCO-ISL EO data diffusion method that leverages the orbit-as-node framework to address these issues.

%% file: sections/8-conclusion.tex
\section{Conclusion}

In this paper, we proposed the OrbitTransit to address GS congestion during delay-tolerant delivery in LSNs. To tackle the complexity and dynamics introduced by frequent GSL updates, we introduced the orbit-as-node framework to simplify edge connectivity without compromising optimality. The proposed traffic diffusion mechanism distributed congested traffic into neighboring orbits, while the contention-avoidant delivery component coordinated task allocation to prevent resource conflicts on satellites, thereby balancing GS loads and ensuring timely delivery.

We believe this paper only scratched the surface of PCO delivery. For future work, we planned to extend OrbitTransit to incorporate cross-layer optimization, overseas delivery, and replication and consistency mechanisms for data center synchronization, among others.

%% file: sections/9-Appendix.tex
\appendix
\section{Appendix}
\label{sec:appendix}

\subsection{Wildfire Monitoring Example}
\label{sec:appendix_wildfire}

\textbf{Adapting to task urgency.} OrbitTransit can be directly integrated into existing EO mission workflows, with wildfire monitoring as one representative example. EO satellites periodically scan fire-prone regions and report data to the task control center (e.g., ESMO) for risk assessment, where routine wildfire products typically tolerate deadlines of around $3$ hours~\cite{NASA_FIRMS_INFORMATION}. Once a high-risk area is identified, the control center can issue follow-up tasks with higher revisit frequency and tighter deadlines, potentially within minutes~\cite{NASA_FIRMS_UltraRealTime}. OrbitTransit then adapts its delivery strategy accordingly: routine data can use PCO to reduce ISL usage and battery consumption, while urgent follow-up observations can re-execute the contention-avoidant delivery algorithm and, if necessary, fall back to ISL-only transmission.

\textbf{Boosting capacity and coverage.} Current NOAA satellites used for wildfire detection still rely on direct broadcast to offload EO data, with a maximum data rate of 25 Mbps at 7812 MHz~\cite{NOAA_JPSS_HighRateData}. This limited downlink capacity may become insufficient as EO tasks continue to increase in scale, sensing quality, and observation frequency. Moreover, existing direct-broadcast receiving stations are primarily maintained by the observation community, which limits their deployment scale and global coverage. By leveraging hybrid PCO-ISL delivery over commercial GS infrastructures, OrbitTransit can substantially improve both delivery capacity and service coverage for wildfire monitoring.

\subsection{PuLP Modeling and Benchmarking}
\label{sec:appendix_pulp}

\textbf{Big-M linearization.} All variables, constraints, and formulations that do not involve exponentiation, logarithmic, or trigonometric functions are encoded in the same manner as described in \S\ref{sec:modeling}. For formulations that include products of two decision variables, such as $x^{j}(s_i^t)$ and $z_{\hat{t}}^k(s_i^t)$ in Eq.~\ref{eq:task_delivery_time}, we linearize these bilinear terms using the big-M method. For simplicity, we omit the task $s_i^t$ and the subscripts and superscripts of $x^{j}(s_i^t)$ and $z_{\hat{t}}^k(s_i^t)$ in the following discussion.

Since both $x$ and $z$ are binary variables, the big-M parameter can be set to $1$. We introduce a new binary auxiliary variable $\alpha$ with the following constraints:
\begin{equation}
\label{algo:alpha_cons}
\begin{aligned}
&\alpha \leq x,\\
&\alpha \leq z,\\
&\alpha \geq x + z - 1.
\end{aligned}
\end{equation}

These constraints ensure that $\alpha = x \cdot z$ holds, thereby eliminating the bilinear term from the formulation. Thus, for each $x \cdot y$ product in Eq.~\ref{eq:task_delivery_time}, we introduce a new auxiliary variable $\alpha$ to replace it, and PuLP must determine the optimal value of $\alpha$ subject to constraint~(\ref{algo:alpha_cons}) as well as the nested constraints on $x$ and $y$ described in \S\ref{sec:modeling}.

\textbf{Linear interpolation for energy model.} The life consumption function in Eq.~\ref{eq:life_consump} includes an exponential term in Euler's number, which cannot be handled directly by linear programming. To linearize it, we isolate the nonlinear part and rewrite Eq.~\ref{eq:life_consump} in the simplified form $f(x) = e^{x}$, omitting the remaining terms since they are linear. We then initialize $m$ breakpoints $\{a_1, \dots, a_m\}$ that uniformly partition the range $[0, 1]$, which matches the domain of a DoD value. Their corresponding function values are precomputed as
\begin{equation}
b_i = f(a_i), \quad \forall i \in \{1, \dots, m\}.
\end{equation}

We then introduce interpolation weights $\lambda_i \in [0, 1]$ whose sum is constrained to $1$:
\begin{equation}
\label{algo:lambda_sum_cons}
\sum_{i \in \{1, \dots, m\}} \lambda_i = 1.
\end{equation}
The linear surrogate $\tilde{f}(x')$ and its input $x'$ are given by
\begin{equation}
\label{algo:input_match_cons}
x' = \sum_{i \in \{1, \dots, m\}} \lambda_i a_i,
\end{equation}
\begin{equation}
\tilde{f}(x') = \sum_{i \in \{1, \dots, m\}} \lambda_i b_i.
\end{equation}
Given any input $x'$ and a set of coefficients $a_i$, there must exist a combination of $\lambda_i$ that satisfies constraint~(\ref{algo:input_match_cons}) due to the convex combination representation of affine functions~\cite{vielma2015mixed}. Therefore, the approximated value of $f(x')$ can be expressed using the same set of $\lambda_i$ and $b_i$. Since the surrogate $\tilde{f}(x')$ is linear, the variables $\lambda_i$ can be jointly solved in PuLP under constraints~\eqref{algo:lambda_sum_cons} and~\eqref{algo:input_match_cons}, where the coefficients $a_i$ and $b_i$ are precomputed from $f(x)=e^{x}$.

\begin{figure}[t]
     \centering
     \includegraphics[width=\linewidth]{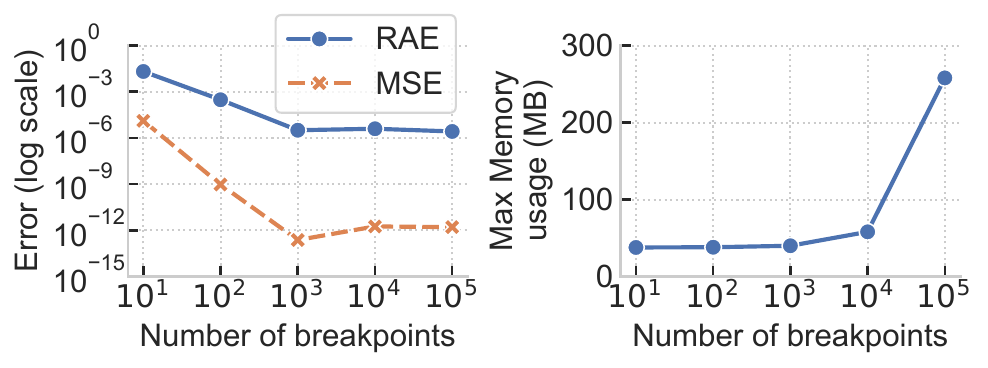}
     \caption{Trade-off between the number of breakpoints, approximation error, and memory usage in the linearized energy model.}
     \label{fig:breakpoints}
\end{figure}

\begin{figure}[t]
     \centering
     \includegraphics[width=\linewidth]{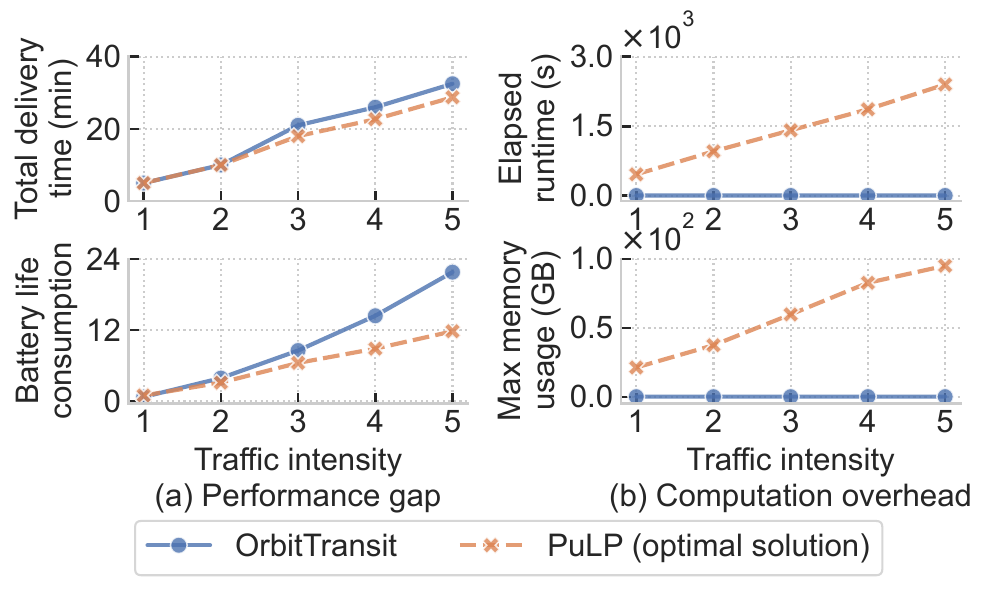}
     \caption{Performance and computational differences between OrbitTransit and the optimal solution.}
     \label{fig:overall_compare_with_pulp}
\end{figure}

\textbf{Hyperparameter setups.} To determine a suitable breakpoint granularity, we uniformly sample points in the range $[0, 1]$ for $1\times10^{3}$ iterations and compute the average relative approximation error (RAE) and mean squared error (MSE) between $f(x)$ and $\tilde{f}(x')$. The results are shown in Figure~\ref{fig:breakpoints}. We observe that $1{,}000$ breakpoints already achieve an RAE of $3.19\times10^{-6}$ and an MSE of $2.37\times10^{-13}$, while finer granularities do not significantly improve accuracy and can even introduce slight degradation due to floating-point precision and PuLP’s solver tolerance, in addition to increasing memory usage. Therefore, we set $m$ to $1{,}000$ in our experiments to balance accuracy and efficiency.

\textbf{Experiment results.} Due to the complexity of the space routing and task placement problem and the large scale of the satellite constellation, the PuLP fails to resolve even the smallest $2$~TB traffic-intensity case, encountering an out-of-memory error despite the $512$~GB of available DRAM in our testbed system. To enable tractable analysis, we scale down the constellation to focus on specific RoI regions rather than the global regions, and further reduce the EO data volume accordingly to maintain the same level of traffic intensity.

The performance gap and computation overheads are shown in Figure~\ref{fig:overall_compare_with_pulp}. The performance of OrbitTransit in terms of delivery time and battery life consumption is comparable to PuLP, with average differences of $2.01$ minutes and $3.62$, respectively. However, PuLP incurs unacceptable computation overhead, as both its elapsed runtime and memory usage grow exponentially with increasing task scale, reaching 2400 seconds and 94~GB, respectively, whereas OrbitTransit maintains a maximum elapsed runtime and memory usage of only 6.37 seconds and 0.21~GB. These results indicate that although OrbitTransit is not theoretically optimal, its performance gap remains acceptable, and its computation overhead is substantially more practical and efficient when deployed on real-world computational units in GSs.

\subsection{Impact of delayed data plane states}
\label{sec:appendix_delay}

\begin{figure}[t]
     \centering
     \includegraphics[width=\linewidth]{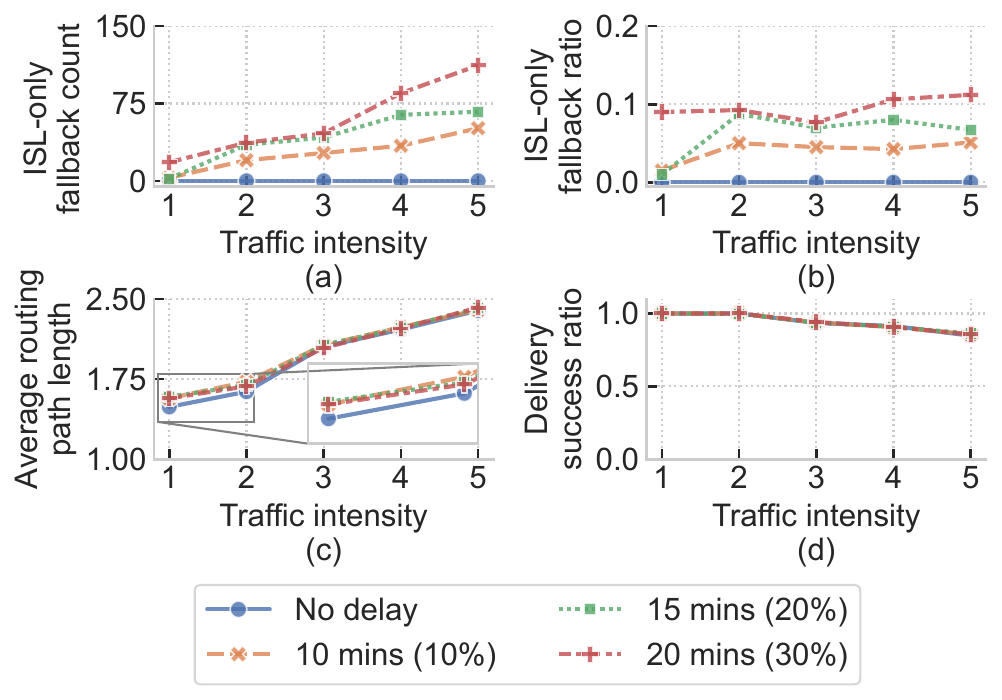}
     \caption{Sensitivity of OrbitTransit to data plane delay. Legend entries X mins (Y\%) denote telemetry delayed by X minutes with probability Y\%.}
     \label{fig:delta_delay_result}
\end{figure}

\textbf{Telemetry delay simulation setup.} To better reflect real-world telemetry delays, we maintain a historical buffer for the data plane states shown in Figure~\ref{fig:framework_overview}. When OrbitTransit queries the load or energy status of a ground station $g_j$ or satellite $sat_k$, it receives a delayed historical state instead of the latest one. After GS selection and routing decisions are made based on this delayed information, the actual delivery is still executed using the real-time states of $g_j$ and $sat_k$. This setup allows us to evaluate how OrbitTransit performs when its control decisions are based on stale data plane observations.

\textbf{Experiment results.} We define three telemetry-delay levels, as shown in the legend of Figure~\ref{fig:delta_delay_result}. Delayed telemetry samples are injected independently at random. As shown in Figure~\ref{fig:delta_delay_result}(a), OrbitTransit triggers more ISL-only fallback as the delay level increases, with the fallback count increasing by $1.21\times$ from the 10 min (10\%) setting to the 20 min (30\%) setting. This is because stale status may cause the scheduled path or destination to become infeasible, in which case OrbitTransit falls back to ISL-only transmission, as described in Section~\S\ref{sec:sys_design}, to ensure successful delivery. To quantify how many tasks are affected by stale states, we define the fallback ratio as the number of tasks that trigger ISL-only fallback divided by the total number of tasks. Figure~\ref{fig:delta_delay_result}(b) shows that this ratio is largely independent of traffic intensity, indicating that the higher fallback count mainly comes from the increased number of tasks. Even under the 20 min (30\%) setting, only $9.54\%$ of tasks are affected, since the orbit-as-node model relies on orbital-level aggregates and is inherently tolerant to stale or partially inaccurate states.

The impact on overall delivery performance is also limited. As shown in Figure~\ref{fig:delta_delay_result}(c), the average routing path length increases by only $2.15\%$ from the no-delay setting to the 20 min (30\%) setting, mainly because the affected $9.54\%$ of tasks fall back to pure-ISL transmission. Meanwhile, the delivery success ratio remains unchanged, since OrbitTransit is a hybrid PCO-ISL system and can adapt to real-time routing conditions by switching to feasible ISL-based delivery when the originally scheduled path becomes infeasible.